\documentclass[printer]{aa}
\usepackage{natbib}
\usepackage{amssymb}
\usepackage{graphicx}
\usepackage{hyperref}
\usepackage{comment}
\usepackage{subcaption}

\graphicspath{{./figures/}}

\def\der{\hbox{d}}

\newcommand{\Msun}{\mbox{$\mathrm{M}_{\odot}$}}

\makeatletter
\newcommand\footnoteref[1]{\protected@xdef\@thefnmark{\ref{#1}}\@footnotemark}
\makeatother

\begin{document}

\title{Disc truncation in embedded star clusters:\\ Dynamical encounters versus face-on accretion}
\titlerunning{Disc truncation in embedded clusters}
\author{T.P.G. Wijnen\inst{1,2}, O.R. Pols\inst{1}, F.I. Pelupessy\inst{2,3}, S. Portegies Zwart\inst{2}}
\authorrunning{T.P.G. Wijnen et al.}
\institute{Department of Astrophysics/IMAPP, Radboud University Nijmegen, P.O. Box 9010, 6500 GL Nijmegen, The Netherlands\\
\email{thomas.wijnen@astro.ru.nl}
\and Leiden Observatory, Leiden University, PO Box 9513, 2300 RA Leiden, The Netherlands
\and CWI, P.O. Box 94079, 1090 GB Amsterdam, The Netherlands
\offprints{T.P.G. Wijnen}
}
\date{Received ..../ Accepted ....}

\abstract{Observations indicate that the dispersal of protoplanetary discs in star clusters occurs on time scales of about 5 Myr. Several processes are thought to be responsible for this disc dispersal. Here we compare two of these processes: Dynamical encounters and interaction with the interstellar medium, which includes face-on accretion and ram pressure stripping. We perform simulations of embedded star clusters with parameterisations for both processes to determine the environment in which either of these processes is dominant. We find that face-on accretion, including ram pressure stripping, is the dominant disc truncation process if the fraction of the total cluster mass in stars is $\lesssim 30\,\%$ regardless of the cluster mass and radius. Dynamical encounters require stellar densities $\gtrsim 10^4$ pc$^{-3}$ combined with a mass fraction in stars of $\approx 90\,\%$ to become the dominant process. Our results show that during the embedded phase of the cluster, the truncation of the discs is dominated by face-on accretion and dynamical encounters become dominant when the intra-cluster gas has been expelled. As a result of face-on accretion, the protoplanetary discs become compact and their surface density increases. In contrast, dynamical encounters lead to discs that are less massive and remain larger.}

\maketitle

\section{Introduction}
The clustered environments in which stars are generally born \citep{lada03} can have a severe impact on the protoplanetary discs of their newborn members. The occurrence frequency of protoplanetary discs in young star clusters is observed to decrease with an e-folding time scale of the order of 5 Myr \citep{cloutier14}. Both internal and external processes influence the dispersal of protoplanetary discs in dense star clusters. Internal processes, such as depletion by accretion onto the central star \citep[e.g.][]{hartmann98-1} and subsequent photoevaporation by the central star \citep[e.g.][]{bally82, shu93, clarke01, alexander06, alexander06-1} are believed to dissipate the discs on time scales $\lesssim 10$ Myr \citep{dullemond06, williams11}. 

Furthermore, observations indicate that external photoevaporation by nearby O stars can reduce the fraction of stars that have a disc by a factor of two \citep[e.g.][]{balog07, guarcello07, guarcello09, fang12, fang13}, although these results can, in part, be explained by sample incompleteness \citep{richert15}. \citet{vicente05} find no obvious correlation between disc sizes and their proximity to OB stars in the Trapezium cluster. The mass-loss rate from the outer edge of the protoplanetary disc due to external photoevaporation can be up to $10^{-8} - 10^{-7}\, \Msun / \mathrm{yr}$ \citep{facchini16}. However, to truncate disc radii to values smaller than 100 AU within a time scale of 10 Myr, ultraviolet fluxes that are at least an order of magnitude higher than typical values in a moderate-sized cluster ($N_* = 300$) are required \citep{adams10}. \citet{mann10} showed that despite the hostile environment around the massive star $\theta^1$ Ori C, the discs in the Orion Nebula cluster (ONC) have similar properties to those observed in isolated low-mass star-forming regions. In this work, we focus on embedded clusters, in which the presence of the pristine gas is expected to absorb radiation and reduce the effect of external photoevaporation.

Apart from (external) photoevaporation, the effects of stellar fly-bys on protoplanetary discs have also been investigated \citep[e.g.][]{clarke93, ostriker94, heller95, kobayashi01}. \citet{scally01} modelled the ONC, using fixed disc radii of 10 and 100 AU, and find that stellar encounters probably do not play a significant role in the destruction of protoplanetary discs. However, \citet{olczak06} also modelled the ONC and their study suggests that up to 15\,\% of the discs may be destroyed by stellar encounters. Moreover, \citet{portegies_zwart16} finds that the observed distribution of disc sizes in the Orion Trapezium cluster can be reproduced by only considering truncation due to dynamical encounters, assuming initial disc radii of 400 AU and an initial fractal distribution for the stars. Both hydrodynamical and gravitational $N$-body studies show that truncation due to dynamical encounters is important in environments with stellar densities $> 10^3\,\mathrm{pc}^{-3}$ \citep{rosotti14, vincke15}. Observational studies also show that the observed size distribution of protoplanetary discs depends on the ambient stellar density \citep{de_juan_ovelar12}. 

In recent studies, \citet{wijnen16, wijnen17} suggest that the ambient gas in embedded star-forming regions can also reduce the sizes of protoplanetary discs. On the one hand, the ram pressure exerted by the interstellar medium (ISM) can strip the outer parts of the protoplanetary disc. On the other hand, the continuous accretion of ISM with little to no azimuthal angular momentum with respect to the disc material, which we call face-on accretion, causes the disc to contract. \citet{wijnen17} provide a parameterisation for the evolution of the disc as a function of the ambient gas density and relative velocity. Here we use this parameterisation in $N$-body simulations with a static gas potential and we compare its effect with that of dynamical encounters. \citet{vincke16} find that disc truncation due to dynamical encounters is most effective in the embedded phase of the cluster. We investigate in which part of the parameter-space each of these two effects is expected to determine the disc-size distribution. We also investigate whether or not,  and how, the two processes affect the disc masses. To allow for a clear comparison, we do not take external photoevaporation into account. As discussed above, this may be a fair assumption to first order.

\section{Methods}

Our simulations are carried out within the \textsc{amuse} framework \citep{portegies_zwart13, pelupessy13}\footnote{\url{http://www.amusecode.org}}. The \textsc{amuse} framework is written in \textsc{python} and provides a user-friendly way to combine different astrophysical packages and software, for example $N$-body integrators and stellar evolution codes. Our setup is practically identical to the setup used by \citet{portegies_zwart16} to model the disc size distribution as a result of dynamical encounters in the Trapezium cluster. We perform $N$-body simulations of star clusters in a static gas potential. Each star is given a parameterised disc and we follow the evolution of the radii and masses of the discs under the influence of face-on accretion and dynamical encounters using parameterisations for both processes. Each of these two processes is considered independently in our simulations. Below, after summarising the method used to resolve dynamical encounters, we describe our implementation of the face-on accretion model of \citet{wijnen17} and the addition of a static gas potential.

We assume the surface density profile of the protoplanetary discs initially follows a power-law, $\Sigma(r) = \Sigma_0(r/r_0)^{-n}$, with $n=1.5$, corresponding to a minimum-mass solar nebula \citep{weidenschilling77, hayashi81}. This allows us to set the inner radius of the disc to 0 AU, instead of a value $\mathcal{O}$(1-10 AU), without realising an unrealistic mass distribution over the disc which would occur for $n=1$. Following \citet{portegies_zwart16}, we assume the initial outer radius of all protoplanetary discs is 400 AU and their mass is equal to 10\,\% of the mass of their host star. We verify that the discs satisfy the \citet{Toomre64} criterion for stability against self-gravity.

\subsection{Resolving dynamical encounters}

We use the same method as described in Sects. 2.1 and 2.2 of \citet{portegies_zwart16} to resolve dynamical encounters. That is, when stars come within a certain encounter radius (initially 0.02 parsec) from each other, the dynamical encounter is resolved by calculating the peri-centre distance, $r_{\rm peri}$, with Kepler's equation (using the kepler module from the \textsc{starlab} package; \citealt{portegies_zwart01}). We calculate the new disc radius, $R'_{\rm disc}$, for a star with mass $M_1$ from the estimate found by \citet{breslau14} for coplanar, pro-grade and parabolic encounters\footnote{Recently, \citet{bhandare16} derived a parameterisation by averaging over all inclinations. Their parameterisation is less restrictive because coplanar, pro-grade encounters have the strongest effect on the disc. We choose the most restrictive prescription to model the strongest influence dynamical encounters can have on protoplanetary discs.}:
\begin{equation}\label{eq:trunc_encounter_application}
R'_{\rm disc} = 0.28 r_{\rm peri} \left(\frac{M_1}{M_2}\right)^{0.32},
\end{equation}  
with $M_2$ being the mass of the other star. This parameterisation has been derived using test particles, that is, it does not consider hydrodynamical forces. Previous work shows that excluding pressure, self-gravity, and viscous effects in simulations of discs in dynamical encounters provides results that are consistent with simulations that include these effects \citep{pfalzner05}. Only for very close encounters, ${R'_{\rm disc} \lesssim 0.2 R_{\rm disc}(t=0)}$, should viscous forces be accounted for. It is questionable whether or not a disc-like structure survives after such a destructive encounter, in which even stellar capture may occur \citep{munoz15}. For disc radii ${\lesssim 0.2 R_{\rm disc}(t=0)}$, the parameterisation may no longer be valid and such small radii must therefore be interpreted with care.

The mass that is lost from the disc during the encounter is calculated using:
\begin{equation}\label{eq:dm_encounter_application}
\Delta M = M_{\rm disc} \frac{R_{\rm disc}^{0.5}-{R'^{0.5}_{\rm disc}}}{R_{\rm disc}^{0.5}},
\end{equation} 
which follows from the assumption that all material beyond $R'_{\rm disc}$ is stripped. The disc of the other star accretes a fraction of the mass that is lost, which we compute as:
\begin{equation}\label{eq:dm_acc_encounter_application}
\Delta M_{\rm acc, 2} = \Delta M \frac{M_2}{M_2+M_1}.
\end{equation}
Eqs.~\ref{eq:trunc_encounter_application}, \ref{eq:dm_encounter_application}, and \ref{eq:dm_acc_encounter_application} are all applied symmetrically in a dynamical encounter.

\subsection{Parameterising face-on accretion}\label{sec:faceontheory_application}

To parameterise face-on accretion of ambient gas onto protoplanetary discs, we use the theoretical model derived in \citet{wijnen17}, which has been shown to adequately describe the evolution of the radius, mass, and surface density profile of the disc. This model is an extension of the thin accretion disc theory, assuming that the mass flux onto the disc is uniform over the surface area of the disc and completely accreted by each surface element. In this model, the disc is considered as consisting of concentric rings. The accretion of ISM with no azimuthal angular momentum decreases the specific angular momentum of each ring, which then contracts to a smaller radius that corresponds to its new specific angular momentum. The differential equations for the radial drift velocity and for the surface density of the disc then obtain an extra ISM accretion term with respect to the thin accretion disc theory. When viscous effects are neglected, these differential equations can be simplified to a scale-free theoretical model. This model depends on a dimensionless parameter $\tau$, which is proportional to the time-integrated mass flux and is defined as: 
\begin{equation} \label{eq:tau_application}
\tau = \frac{5}{\Sigma_0(r_0)} \int_0^t \rho(t') v(t') \cos{i(t')} \,\der t',
\end{equation}
where $\Sigma_0(r_0)$ is the initial surface density of the disc at an arbitrary but fixed scaling radius $r_0$. The mass flux onto the disc is determined by the density of the ambient medium $\rho$ and velocity relative to this medium $v$. The factor $\cos{i(t)}$, where $i$ is the inclination between the angular momentum vector of the disc and the flow direction of the ISM, accounts for the situation in which the disc is not aligned perpendicular to the mass flux.
Assuming an initial surface density profile, $\Sigma_0(r/r_0)^{-n}$, the surface density of the disc at any moment in time is then given by:
\begin{equation}\label{eq:sigma_application}
\Sigma(r, \tau) = \Sigma_0 \left[\left(\frac{r}{r_0}\right)^{-n} + \tau \right].
\end{equation}
Using the parameter $\tau$, the differential equation for any radius within the disc, $r(t)$, can be expressed in terms of a dimensionless radius $y(\tau)$ via $r(t) = y(\tau) r_0 $. The differential equation for $y(\tau)$ can be written as \citep[Eq. 16 from][]{wijnen17}:
\begin{equation} \label{eq:dydtau_application}
\frac{\der y}{\der\tau} = - \frac{2}{5} \, \frac{y}{y^{-n} + \tau}.
\end{equation}
We calculate the solution to Eq.~\ref{eq:dydtau_application} for an initial value $y_0 = 1$ and call it $Y_1(\tau)$. The evolution of any other annulus in the disc with initial value $y_0 = r(0)/r_0$ can then be derived from:
\begin{equation}\label{eq:Ytau_application}
y(\tau) = y_0 Y_1({y_0}^n\tau),
\end{equation}
where we have taken advantage of the self-similarity of the solutions. Rather than integrating Eq.~\ref{eq:dydtau_application} for every arbitrary starting radius $y_0$, we obtain its evolution from $Y_1(\tau)$ using Eq.~\ref{eq:Ytau_application}. In our simulations, we follow the evolution of the outer radius of each disc, $R_{\rm disc}(t)$, via its dimensionless equivalent $y_{\rm disc}(\tau)$. For each disc, the evolution of $y_{\rm disc}(\tau)$ is determined by its initial surface density $\Sigma_0(r_0)$ and its time-integrated mass flux. The solutions to Eq.~\ref{eq:Ytau_application} therefore describe the radius of any disc, at any time, via its individual parameter $\tau$. In our simulations, the integration in Eq.~\ref{eq:tau_application} is replaced by a summation over all time steps. At each time step we use the velocity of the star and the density at its position. The time step we use is defined in Sect.~\ref{sec:bridge_application}. 

The disc may also be truncated by the ram pressure, $P_{\rm ram} = \rho v^2$, of the ambient medium. We compute the truncation radius, $R_{\rm tr}$, by solving the equation derived in Sect. 2.1 of \citet{wijnen17} but  modified to account for the perpendicular velocity component of the flow:
\begin{equation}\label{eq:Rtrunc_application}
R^2_{\rm tr}(t) = \frac{GM_*\Sigma(R_{\rm tr}(t), \tau)}{2 \pi \rho(t) v^2(t) \cos{i(t)}}.
\end{equation}
At every time step we check if the truncation radius is smaller than the disc radius using the current parameters and solving for the radius with the Newton-Raphson method. \citet{namouni07} derived an exact solution for the truncation radius for the purely dynamical problem, in which a particle is ejected from its orbit by a constant force. This solution differs by a factor of $\approx 2$ in the denominator of Eq.~\ref{eq:Rtrunc_application}. Neither solution takes the gas-dynamical and viscous effects in the disc into account. We have verified that Eq.~\ref{eq:Rtrunc_application} provides a better approximation of the results in the hydrodynamical simulations of \citet{wijnen17} and we therefore choose to use this approximation.

Let us suppose a disc is truncated due to ram pressure stripping at a certain moment, $t_{\rm tr}$, corresponding to $\tau = \tau_{\rm tr}$ via Eq.~\ref{eq:tau_application}; after which its new radius is $R_{\rm tr} = \hat{y}(\tau_{\rm tr}) r_0$. Then, according to Eq.~\ref{eq:Ytau_application}, we can write:
\begin{equation}\label{eq:ytr_application}
\hat{y}(\tau_{\rm tr}) =  \hat{y}_0 Y_1({\hat{y}_0}^n\tau_{\rm tr}).
\end{equation}
In order to compute its further evolution with Eq.~\ref{eq:Ytau_application}, we must solve Eq.~\ref{eq:ytr_application} for the unknown new initial value $\hat{y}_0$. We do this in the following manner: We define an inverse function, $X_s(\tau)$, that for any $\tau$ returns the initial value $y'_0$ for which $y'(\tau) = s$. We have calculated these initial values for $s=1$ by solving the differential Eq.~\ref{eq:dydtau_application} up to the moment $y'(\tau) = 1$ for a large number of initial values $y'_0$. This yields unique pairs of $\tau$ and corresponding $y'_0$ values. These pairs form an interpolation table, which is our function $X_1(\tau)$. We can rewrite Eq.~\ref{eq:Ytau_application} in terms of $X_1(\tau)$ by replacing $y(\tau)$ with $y'(\tau)$ for which $y'(\tau) = 1$ and $X_1(\tau) = y'_0$:
\begin{equation}\label{eq:XY_application}
1 = X_1(\tau) Y_1\left[{X_1}^n(\tau)\tau\right],
\end{equation}
which is true for any $\tau$. Using $\hat{y}(\tau_{\rm tr}) = R_{\rm tr}/r_0$, we can also rewrite Eq.~\ref{eq:ytr_application} to obtain:
\begin{eqnarray}\label{eq:Yt_application}
1& = & \frac{\hat{y}_0}{\hat{y}(\tau_{\rm tr})} Y_1({\hat{y}_0}^n\tau_{\rm tr}) \\\nonumber
& = & \hat{y}_0\frac{ r_0}{R_{\rm tr}} Y_1({\hat{y}_0}^n\tau_{\rm tr}) \\\nonumber
& = & \hat{y}_0 \frac{r_0}{R_{\rm tr}} Y_1\left[{\hat{y}_0}^n \left(\frac{r_0}{R_{\rm tr}}\right)^n\tau_{\rm tr} \left(\frac{R_{\rm tr}}{r_0}\right)^n \right].
\end{eqnarray}
Eqs.~\ref{eq:XY_application} and \ref{eq:Yt_application} can only be equal if $X_1(\tau) = \hat{y}_0 r_0/R_{\rm tr}$ and $\tau = \tau_{\rm tr} (R_{\rm tr}/r_0)^n$. Therefore, after truncation, we can obtain the new initial $\hat{y}_0$ from the truncated disc radius $R_{\rm tr}$ and $\tau_{\rm tr}$ via:
\begin{equation}\label{eq:newy0_application}
\hat{y}_0 = \frac{R_{\rm tr}}{r_0} X_1\left[ \tau_{\rm tr} \left(\frac{R_{\rm tr}}{r_0}\right)^n \right].
\end{equation}
Thus when ram pressure truncates the disc, we use Eq.~\ref{eq:newy0_application} to calculate the new initial value $y_0$ and then use Eq.~\ref{eq:Ytau_application} to compute the further evolution of the disc radius. If the disc is truncated by ram pressure stripping, we also assume all material beyond the truncation radius is removed from the disc as was done in \citet{wijnen17} in their comparison with hydrodynamical simulations. 

For face-on accretion, we calculate the mass of the disc at any moment by integrating over the surface density profile (Eq.~\ref{eq:sigma_application}):
\begin{equation}\label{eq:Mdisc_application}
M_{\rm disc} (\tau) =  2\pi r_0^2 \Sigma_0 \left( \frac{1}{2-n} y^{2-n} _{\rm disc}(\tau)+ \frac{1}{2} y^2_{\rm disc}(\tau) \tau \right),
\end{equation}
 \citep[cf. Eq. 18 from][]{wijnen17}. We give the discs a random orientation vector that represents the angular momentum vector of the disc. At every time step, we calculate the absolute value of the dot product between this orientation vector and the velocity vector to obtain $\cos{i}$, necessary in Eq.~\ref{eq:tau_application}. We take the absolute value because it does not matter whether the disc is rotating prograde or retrograde. Wijnen et al. (2017b, submitted) have shown that when the rotation axis of the disc is inclined with respect to the velocity vector, the two vectors tend to align as the disc experiences a net torque from the flow (even if the disc is initially symmetric). The mechanism that causes the tilting of the disc is known but the dissipative processes that determine the time scale of the process are not fully understood. The time scale estimate derived by Wijnen et al. (2017b) for this process suggests that a change in the inclination generally takes at least $10^5$ yr for the conditions encountered in our simulations, except in the simulations with the most dense initial conditions. Since this is much longer than the crossing time of the cluster, we do not take this tilt process into account.

The model described above is only valid if the accretion of ISM is sufficiently rapid that viscous effects can be neglected; that is, $\tau_{\dot{m}} < \tau_{\nu}$, where $\tau_{\dot{m}}$ is the accretion time scale and $\tau_{\nu}$ the viscous time scale \citep[see Sect. A.2 of][]{wijnen17}. This is valid as long as:
\begin{equation} \label{eq:tau-limit_application}
\tau \lesssim \left[ \frac{\tau_{\nu}(R_{\rm disc, 0})}{\tau_{\dot{m}}(R_{\rm disc, 0})} \right]^{5/6}, 
\end{equation}
where $R_{\rm disc, 0}=y_0 r_0$ is the initial disc radius (to which the disc was truncated in case of ram pressure stripping). We define the accretion time scale as \citep[cf. Eq. 8 in][]{wijnen17}:
\begin{equation}\label{eq:taum_param}
\tau_{\dot{m}} (r) = \frac{\Sigma(r)}{5\rho v \cos{i}} = \frac{\Sigma(r) t}{\Sigma_0(r_0) \tau}.
\end{equation}
In the last equality, we have taken the time-averaged value of $\rho v \cos{i}$, using Eq.~\ref{eq:tau_application}, instead of the instantaneous value. For the viscous time scale, we use Eq. 7 from \citet{wijnen17}, adopting the same viscosity parameter $\alpha = 0.01$ and sound speed $c_s = 0.3$ km/s as in that study. At the end of each simulation we check whether every disc meets the condition in Eq.~\ref{eq:tau-limit_application}, which is the case in all our simulations.  

Over long time scales, the radius and mass of the disc scale as simple power-law functions of $\tau$ \citep{wijnen17}. For  $\tau > y_{\rm disc}^{-n}$, the disc radius approaches $R_{\rm disc} \propto \tau^{-2/5}$, as follows from Eq. ~\ref{eq:dydtau_application}. On the same time scale, the mass of the disc is dominated by the second term inside the brackets in Eq.~\ref{eq:Mdisc_application} and $M_{\rm disc} \propto \tau^{1/5}$. We use these relations in the discussion of our results.

\subsection{Gas density distribution}

We assume the gas in the cluster follows a Plummer profile \citep{plummer11}, which is characterised by a total mass $M_{\rm gas}$ and a characteristic radius $R_{\rm Plummer}$. This distribution allows us to model the initial distribution of the stars and gas consistently: the initial positions of the stars are distributed following the same Plummer profile. A different initial stellar distribution, for example a fractal distribution, cannot be combined with a consistent analytic (equilibrium) gas distribution. On the other hand, modelling the gas with a hydrodynamics code makes the simulations computationally expensive and prevents us from exploring a large parameter space. In this work we explore the regimes in which the process of dynamical encounters and face-on accretion are relevant. In future studies, specific parts of the parameter space can be explored in more detail.

As in \citet{portegies_zwart16} we simulate our simplified star-forming region for 1 Myr, which is motivated by observations of the ONC. The mean stellar age of the ONC is $< 1$ Myr and $\geq$ 80\,\% of the stars in the Trapezium cluster have an inferred age less than 1 Myr  \citep{prosser94, hillenbrand97}. Moreover, young clusters are observed to be embedded during the first 1-3 Myr of their evolution \citep{lada03, portegies_zwart10}. A simulation time of 1 Myr allows us to keep the density distribution time-independent, which reduces the integration error.

\subsection{Initial stellar conditions}

The stars are randomly drawn from a \citet{kroupa01} initial mass function between 0.01 and 100 $\Msun$; the mean mass of this distribution is $0.4\,\Msun$. The mass of the protoplanetary disc ($0.1\,M_*$) is added to the mass of the star in the N-body integrator, which is the fourth-order \textsc{hermite} N-body code \textsc{ph4} (McMillan, publicly available in \textsc{amuse}) and for which we assume a time-step parameter $\eta = 0.01$ and a softening of 100 AU. The initial positions of the stars are assigned according to a Plummer distribution that follows the gas distribution. This simplified assumption may not be correct as studies have shown that the star formation efficiency increases with gas density \citep[e.g.][]{burkert13}. This effect can be partly taken into account by comparing simulations with different fractions of the total cluster mass in stars. We adopt four different values of this fraction, given in Sect.~\ref{sec:parameters_application}. The highest value of $\approx 90\,\%$ could represent either an extreme case of the aforementioned scaling of the star formation efficiency with gas density or a star-forming region at the end of its embedded phase. The cluster is set up without primordial mass segregation. The Plummer profile may underestimate both the stellar and gas density in the centre of the cluster but we show in the results section that as long as both are underestimated consistently our results still hold.
\subsection{Coupling the N-body integrator to the gas potential}\label{sec:bridge_application}

The N-body integrator and analytic potential are coupled using the \textsc{AMUSE} Bridge integrator \citep{fujii07, pelupessy13}, which uses Hamiltonian splitting and a second order leapfrog integration scheme. We choose a Bridge time step of $R_{\rm plummer}/100 v_{\rm esc}$, where we have assumed $v_{\rm esc} = \sqrt{ 2 G M_{\rm tot}/R_{\rm plummer}}$ with $M_{\rm tot}$ the total cluster mass. This time step is small enough to ensure that the amount of swept-up ISM in one time step never exceeds $0.01 M_{\rm disc}$. The energy conservation in the Bridge system is better than a few parts in $10^5$, which is sufficient for a reliable result \citep{portegies_zwart14}.


\subsection{Varying model parameters}\label{sec:parameters_application}

Our simulations are defined by four parameters: The total cluster mass, $M_{\rm tot}$, the virial radius of the cluster $R_{\rm cluster}$, within which 64\,\% of $M_{\rm tot}$ is contained and which is related to the characteristic Plummer radius via $R_{\rm cluster}= (16/3\pi) R_{\rm Plummer}$, the number of stars, $N_*$, and the virial ratio $Q_{\rm vir}$, defined as $Q_{\rm vir} = 0.5 E_{\rm kin}/E_{\rm kin, vir}$, where $E_{\rm kin, vir}$ is the kinetic energy of the stellar component if the stars are in virial equilibrium with the total cluster mass distribution. We checked that in case of virial equilibrium, the radial distribution of the stars remains roughly constant during the simulation. For a given value of $M_{\rm tot}$ we choose four fixed values for $N_*$ such that the fraction of mass in stars (a proxy for the star formation efficiency) is approximately 3, 10, 30 and 90\,\%. We calculate the total stellar mass, $M_{\rm stars}$, and the total gas mass is then given by $M_{\rm gas} = M_{\rm tot} - M_{\rm stars}$. The realisation of the initial stellar mass function determines $M_{\rm stars}$ and the remaining gas mass therefore varies for a given $N_*$. Each of these simulations is performed with a virial ratio of $Q_{\rm vir} = 0.1, 0.5$ and 1.0, representing subvirial, virial, and supervirial initial conditions respectively. Moreover we perform each simulation ten times with a different random seed to reduce the Poisson noise in the final results. We choose small cluster radii such that our simulations can be considered as a local substructure within a larger star-forming region. Our simulated clusters are on the dense end of what is generally observed. In the results section it becomes clear that the most important implications of our simulations can be easily extended to clusters that are less dense. We address this in Discussion Sect.~\ref{sec:dis_observations_application}. The initial conditions are listed in Table~\ref{tb:parameters_application}, which also provides labels for several simulations that we discuss in detail. Two of our sets of initial conditions correspond to the early and late evolution of the Trapezium cluster, respectively, and have been labelled according to the assumed mass fraction in stars.

\begin{table*}
\begin{center}
\caption{Initial conditions for our simulations, which are defined by four parameters: The total mass of the cluster $M_{\rm tot}$, the virial radius of the cluster, $R_{\rm cluster}$, the number of stars, $N_*$, which determines the fraction of mass in stars, and the virial ratio, $Q_{\rm vir}$. We labelled several simulations for ease of reference.}
\label{tb:parameters_application}
\begin{tabular}{lrrr}
\hline
$\mathbf{M_{\rm tot}\,[\Msun]}$ &$\mathbf{R_{\rm cluster}\,[\mathrm{pc}]}$&$\mathbf{N_*}$&$\mathbf{Q_{\rm vir}}$\\
\hline
300&[0.1, 0.25, 0.5]&[20, 70, 205, 610]&[0.1, 0.5, 1.0]\\
1000&[0.1, 0.25, 0.5]&[70, 230, 680, 2045]&[0.1, 0.5, 1.0]\\
3000&[0.1, 0.25, 0.5]&[205, 680, 2045, 6135]&[0.1, 0.5, 1.0]\\
\hline
\end{tabular}
\end{center}
\vspace{4mm}
\begin{center}
\begin{tabular}{lrrrr}
\hline
\textbf{Model}&$\mathbf{M_{\rm tot}\,[\Msun]}$ &$\mathbf{R_{\rm cluster}\,[\mathrm{pc}]}$&$\mathbf{N_*}$&$\mathbf{Q_{\rm vir}}$\\
\hline
Low-mass (LM)&300&[0.1, 0.25, 0.5]&610&0.5\\
Intermediate mass (IM)&1000&0.25&[70, 230, 680, 2045]&0.5\\
High-mass (HM)&3000&[0.1, 0.25, 0.5]&2045&0.5\\
Trap30&1000&0.25&680&0.5\\
Trap90&300&0.25&610&0.5\\

\hline
\end{tabular}
\end{center}
\end{table*}

\section{Results}\label{sec:res_application}

When we refer to the disc radius and disc mass resulting from face-on accretion, we mean the combined effect of face-on accretion and ram pressure stripping.

\subsection{Disc radii}\label{sec:discradii_application}

In Sects.~\ref{sec:res_radiusdistribution_application} and \ref{sec:res_case_application} we discuss the results for the disc radius in the set of IM simulations in detail. The simulation with $N_* = 680$ and an approximate mass fraction in stars of 30\,\% can be considered representative for the early evolution of the Trapezium cluster \citep{lada03} and we therefore label it Trap30. All four simulations have a stellar density $\gtrsim 10^3$ pc$^{-3}$ meaning that truncation due to dynamical encounters should be relevant. 

\subsubsection{Characterising the disc radius distribution}\label{sec:res_radiusdistribution_application}

\begin{figure*}[ht]
    \begin{subfigure}[b]{0.5\textwidth}
        \centering
        \includegraphics[width=\textwidth]{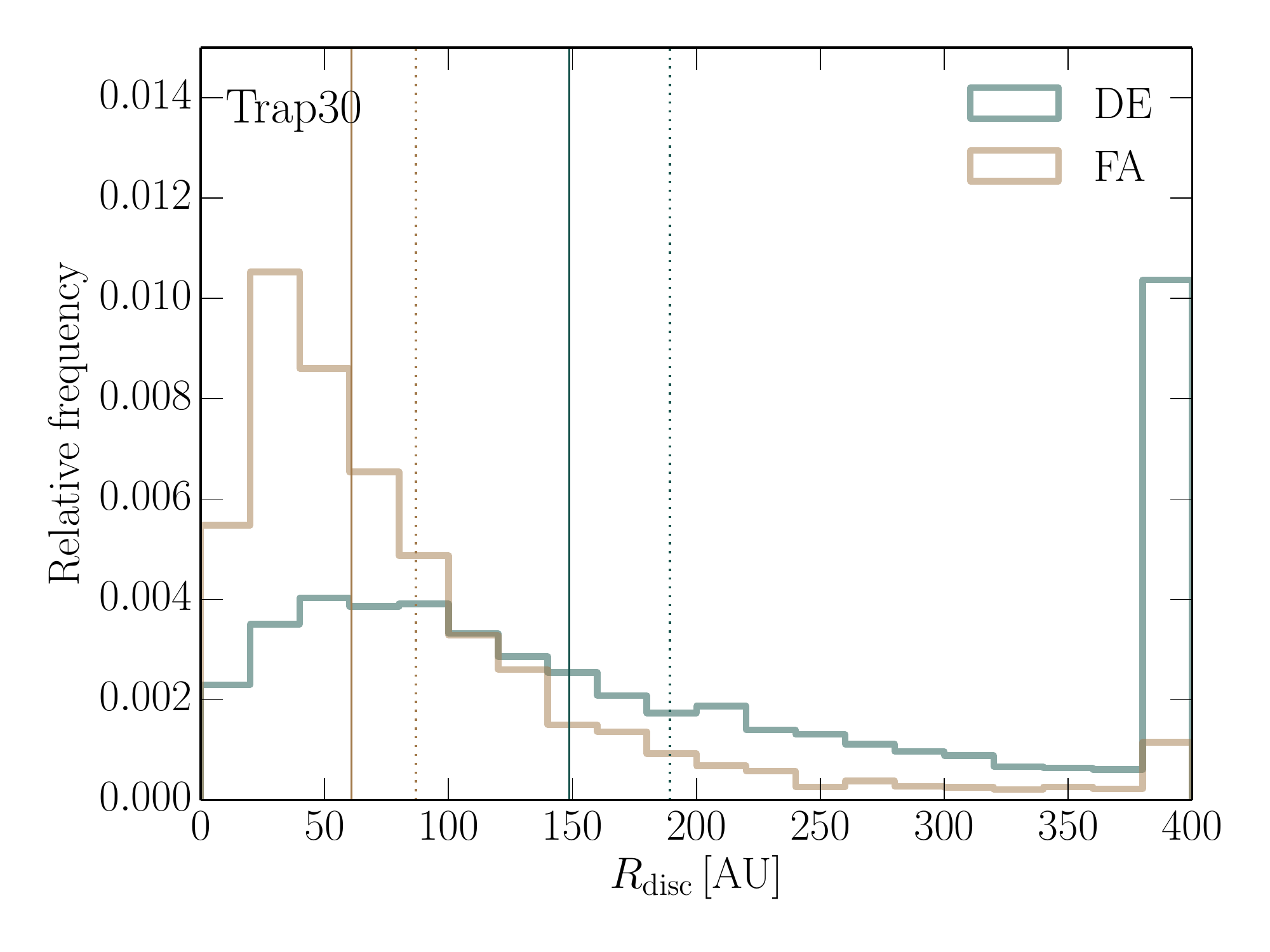}
        \subcaption{}
        \label{fig:hist_application}
    \end{subfigure}
    \begin{subfigure}[b]{0.5\textwidth}
        \centering
        \includegraphics[width=\textwidth]{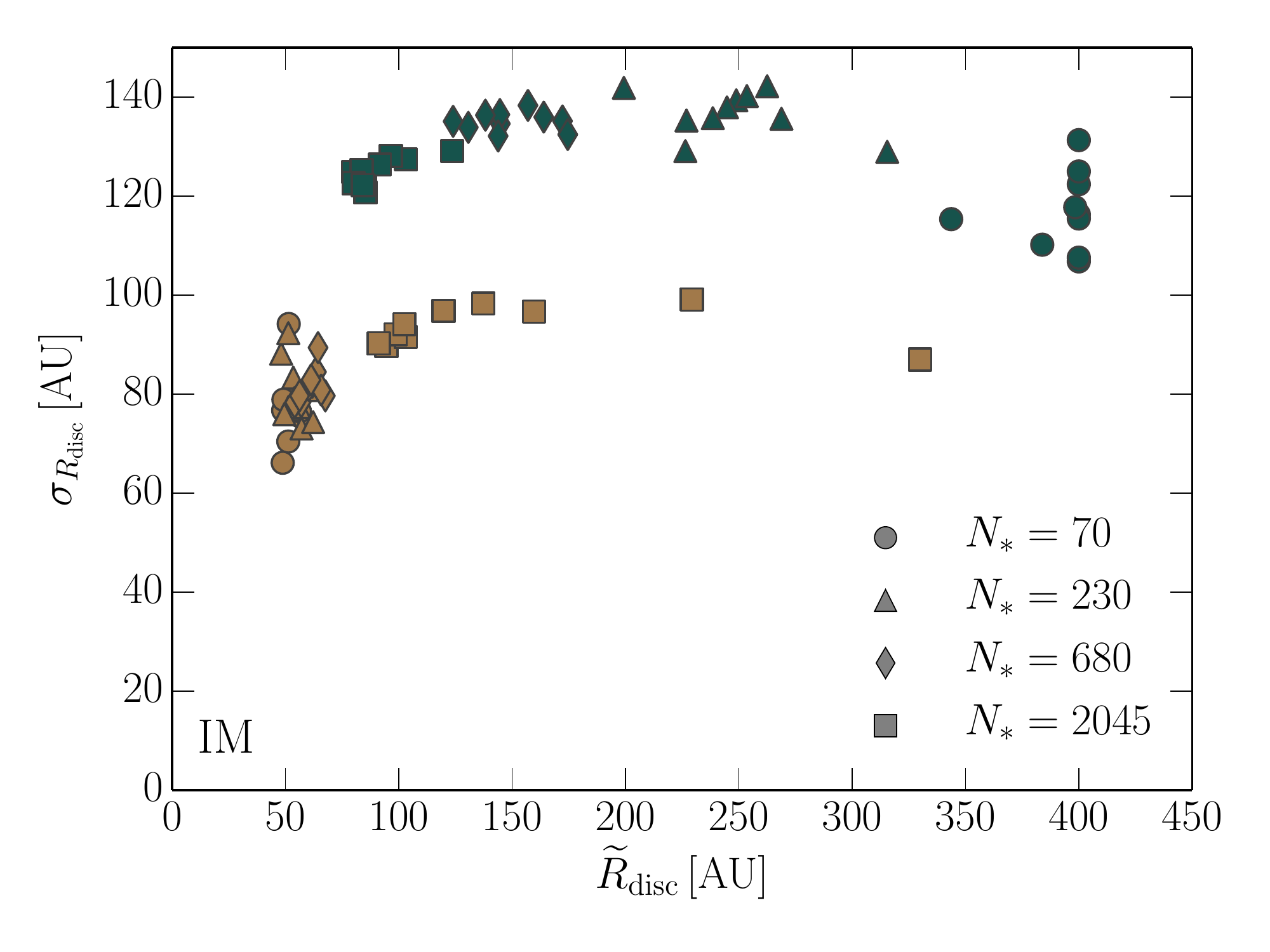}
        \subcaption{}
        \label{fig:median_sigma_application}
    \end{subfigure}
    \caption{\textbf{\emph{(a):}} Relative distribution of disc radii at the end of the simulations (1 Myr) for our Trap30 simulations. Brown corresponds to face-on accretion (FA), including ram pressure stripping, and green to truncation due to dynamical encounters (DE). The distribution is based on all the disc radii in the ten simulations we have performed with these parameters. The solid and dotted vertical lines in the corresponding colour indicate the median and mean disc radius, respectively.\textbf{\emph{(b):}} The standard deviation of the radius distribution versus the median disc radius for the IM simulations. The colours correspond to the same process. Squares indicate the simulations with $N_* = 2045$, diamonds the simulations with $N_* = 680$, triangles the simulations with $N_* = 230$ and circles the simulations with $N_* = 70$.}\label{fig:characterise_radius_application}
\end{figure*}

Because we compare the effect of dynamical encounters and face-on accretion for a large parameter space of initial conditions, it is convenient to characterise the resulting disc radius distribution by, preferably, one value. In Fig.~\ref{fig:hist_application} we show the distribution of disc radii for the Trap30 simulations. For dynamical encounters the distribution is bi-modal, because the process is stochastic and 20\,\% of the stars have not experienced a fly-by that was close enough for truncation. This fraction increases to 56\,\% in the IM simulation with $N_*= 70$. The stars that do not experience a dynamical encounter reside in the outskirts of the cluster, where the stellar density, and hence the number of encounters, is low. Since the distribution of disc radii resulting from dynamical encounters is bi-modal, we choose to use the median disc radius, $\widetilde{R}_{\rm disc}$, instead of the mean disc radius, as a characteristic of the distribution. This provides a better indicator for the influence of dynamical encounters when few discs experience truncation due to a fly-by. For face-on accretion, only a small fraction of the stars - less than 6\,\% in the IM simulations discussed in this section - have a disc radius larger than 380 AU. This fraction does not depend on the number of stars in these simulations. The stars with radii $\ge 380$ AU reside at large distances from the cluster centre, where not only the stellar density but also the ambient gas density and velocities are low. Fig.~\ref{fig:median_sigma_application} shows the standard deviation of the distributions as a function of their median radius. It shows that for each process the standard deviation does not depend strongly on the number of stars. The standard deviation is larger for dynamical encounters due to the skewed distribution of truncated discs. The median disc radius for face-on accretion varies strongly between the ten IM simulations with $N_* = 2045$. The large number of stars drawn from the initial mass function in these simulations leads to large fluctuations in the resulting gas density. We discuss this in the following section. 

Based on the results illustrated by Figs.~\ref{fig:hist_application} and \ref{fig:median_sigma_application}, we conclude that the median disc radius is a good indicator for the effect of each process on the disc-size distribution. For dynamical encounters, the median disc radius does not depend on the initial disc radii, that is, the tail of the distribution. We therefore use the median disc radius in the remainder of this work to quantify the effects of the two processes.

\subsubsection{Comparing stellar mass fractions}\label{sec:res_case_application}

\begin{figure*}[ht]
    \begin{subfigure}[b]{0.5\textwidth}
        \centering
        \includegraphics[width=\textwidth]{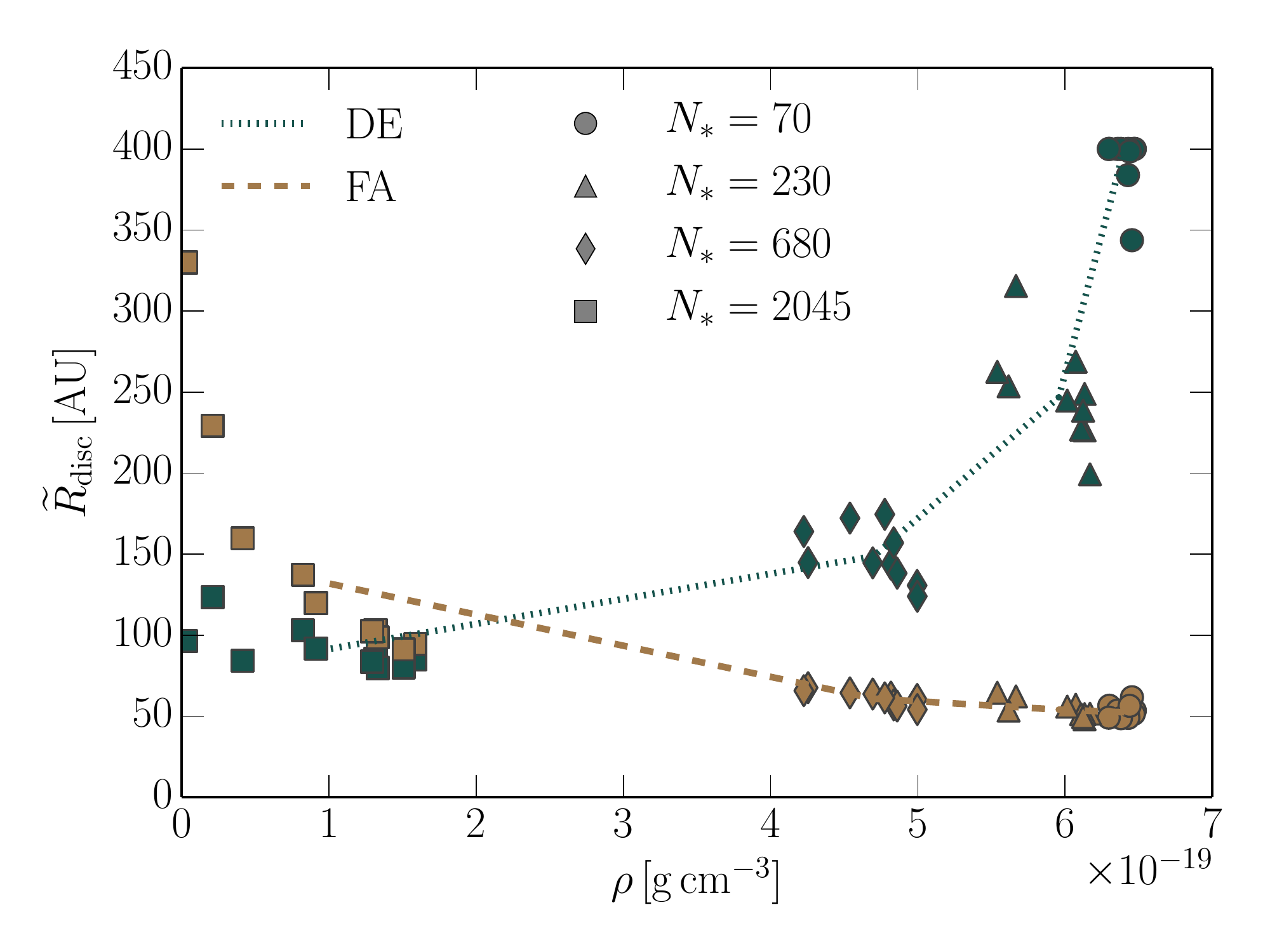}
        \subcaption{}
        \label{fig:case_density_application}
    \end{subfigure}
    \begin{subfigure}[b]{0.5\textwidth}
        \centering
        \includegraphics[width=\textwidth]{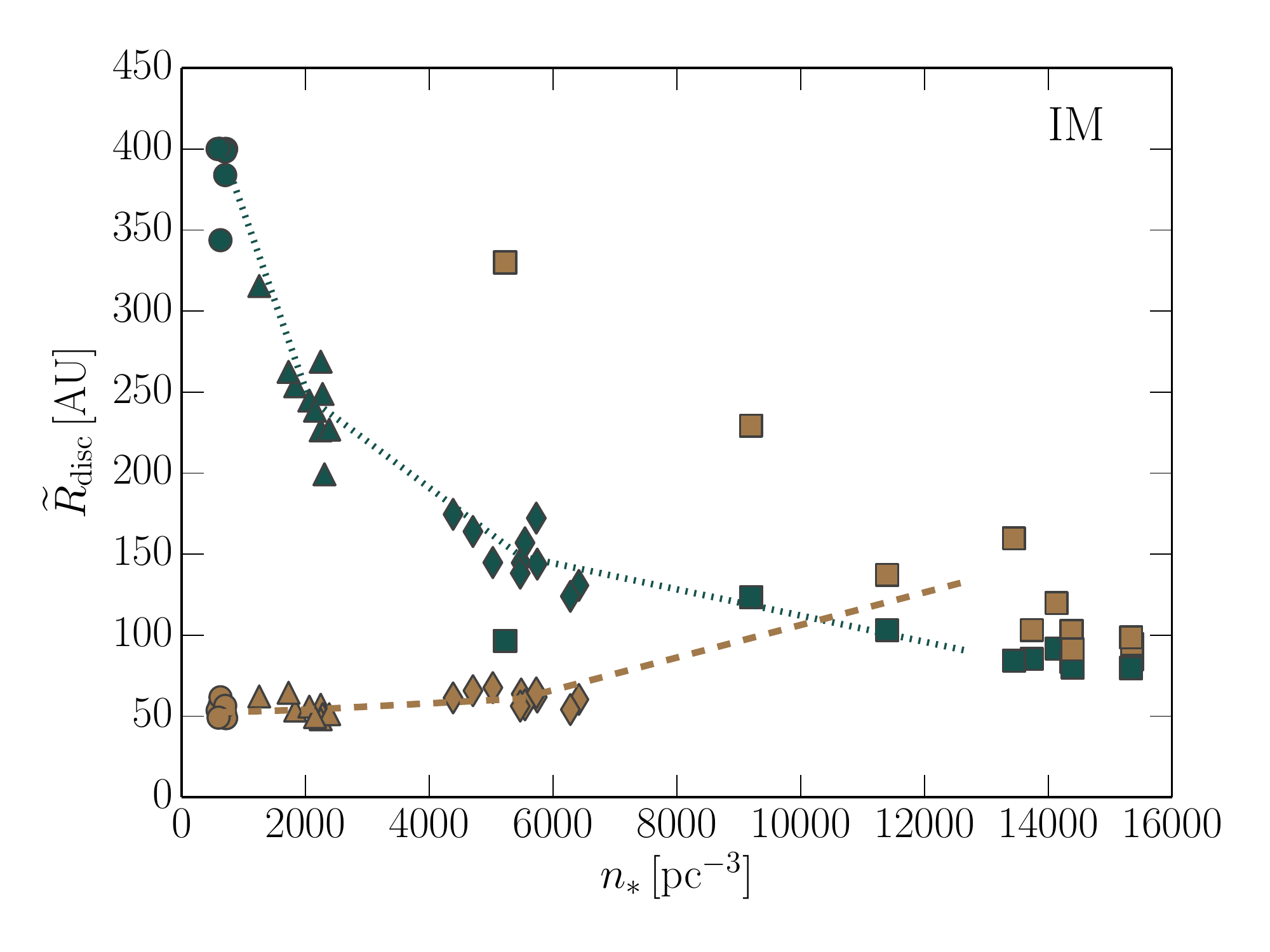}
        \subcaption{}
        \label{fig:case_stellar_density_application}
    \end{subfigure}
    \caption{\textbf{\emph{(a):}} The median disc radius at the end of the simulation plotted against the average gas density within the cluster radius in the IM simulations. The colours and symbols correspond to the same process and number of stars as in Fig.~\ref{fig:characterise_radius_application}. The lines connect the averages of the ten simulations with the same parameters.\textbf{\emph{(b):}} Same as in the left figure, but here we plot the median disc radius against the stellar density. The legend applies to both panels.}\label{fig:case_application}
\end{figure*}

Because the total cluster mass is set, the main difference between simulations with the same initial conditions but different stellar mass fractions is the amount of mass in gas and in stars. The velocity dispersion of the stars isa determined by the total cluster mass and it is therefore practically the same for the IM simulations discussed in this section. Furthermore, the distribution of stars throughout the cluster also remains roughly the same because we assumed the initial cluster to be in virial equilibrium. The discriminating parameters between the simulations with different numbers of stars are the gas and stellar density. We therefore plot the mean disc radius versus these two quantities in Figs.~\ref{fig:case_density_application} and \ref{fig:case_stellar_density_application}, respectively. We determine the stellar density by counting all stars within the cluster radius of 0.25 pc at the end of the simulation and calculate the average gas density within the cluster radius analytically. By definition, the gas density decreases as the number of stars increases. There are some notable outliers in Fig.~\ref{fig:case_application}, particularly for the highest number of stars. These outliers are caused by the sampling of the initial stellar mass function with a fixed number of stars, that continues beyond the intended $M_{\rm stars}$ if $N_*$ is not yet reached. The gas that remains to form stars varies between 5 and 240\,$\Msun$ among simulations that have the same initial conditions otherwise. 

The median disc radius resulting from face-on accretion decreases with increasing gas density. This is expected theoretically because, as discussed at the end of Sect.~\ref{sec:faceontheory_application}, the long-term evolution of an individual disc radius scales as $R_{\rm disc} \propto \tau^{-2/5}$. Of the quantities that determine the value of $\tau$ in Eq.~\ref{eq:tau_application}, the typical velocities and surface densities are similar in each simulation. Averaged over the whole population, only the mean gas density varies between IM simulations with different $N_*$, so that we expect the approximate relation $\widetilde{R}_{\rm disc} \propto \rho^{-2/5}$. In Fig.~\ref{fig:case_density_application} the curve of the median disc radii resulting from face-on accretion follows a slightly steeper trend with density, approximately $\widetilde{R}_{\rm disc} \propto \rho^{-0.5}$. The difference is due to ram pressure stripping becoming more prominent at higher gas densities and causes additional truncation. 

On the other hand, when only dynamical encounters are accounted for, the median disc radius increases strongly with the ambient gas density, that is, with decreasing stellar density as can be seen in Fig.~\ref{fig:case_stellar_density_application}. In simulations with the same total mass and radius, the change in stellar mass fraction affects the number of encounters only via the stellar density, because the velocity dispersion remains almost the same. At a stellar density of 1000 pc$^{-3}$ the effect of dynamical encounters is minor, $\widetilde{R}_{\rm disc} = 400\,$AU, and at a stellar density of $6 \times 10^3$ pc$^{-3}$ the effect of face-on accretion on the disc sizes is still larger than the effect of dynamical encounters. Only for a stellar mass fraction of 90\,\% and corresponding stellar density $> 10^4\,\mathrm{pc^{-3}}$, are dynamical encounters the dominant disc-truncation process. Even at the lowest gas densities, face-on accretion reduces the averaged median disc radius substantially, to 133 AU. This average is determined by three outliers and the rest of the simulations with the same initial conditions indicate that the effect of both processes is comparable. Due to their low gas-mass fractions, these three outlying clusters are more dynamically evolved than the other seven clusters with the same parameters, as can be inferred from the low stellar density within the cluster radius at the end of the simulation. 

These results show that the effect of dynamical encounters relative to face-on accretion becomes important once the gas is (partly) expelled or star formation is very efficient.

\subsubsection{Disc radii as a function of cluster radius and mass}\label{sec:res_radii_parameterspace_application}

\begin{figure*}[ht]
    \begin{subfigure}[b]{0.5\textwidth}
        \centering
        \includegraphics[width=\textwidth]{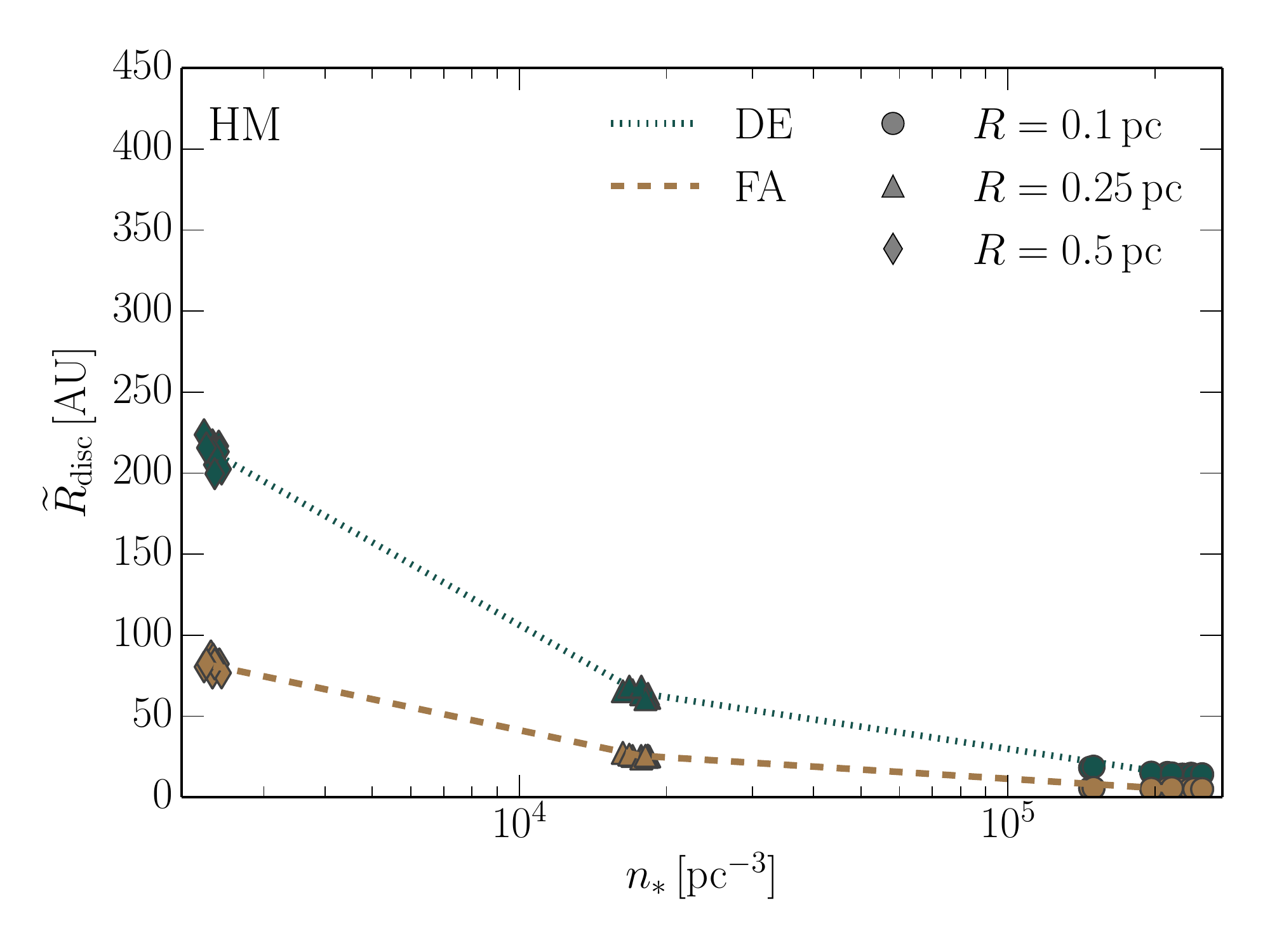}
        \subcaption{}
        \label{fig:3000_stellar_density_application}
    \end{subfigure}
    \begin{subfigure}[b]{0.5\textwidth}
        \centering
        \includegraphics[width=\textwidth]{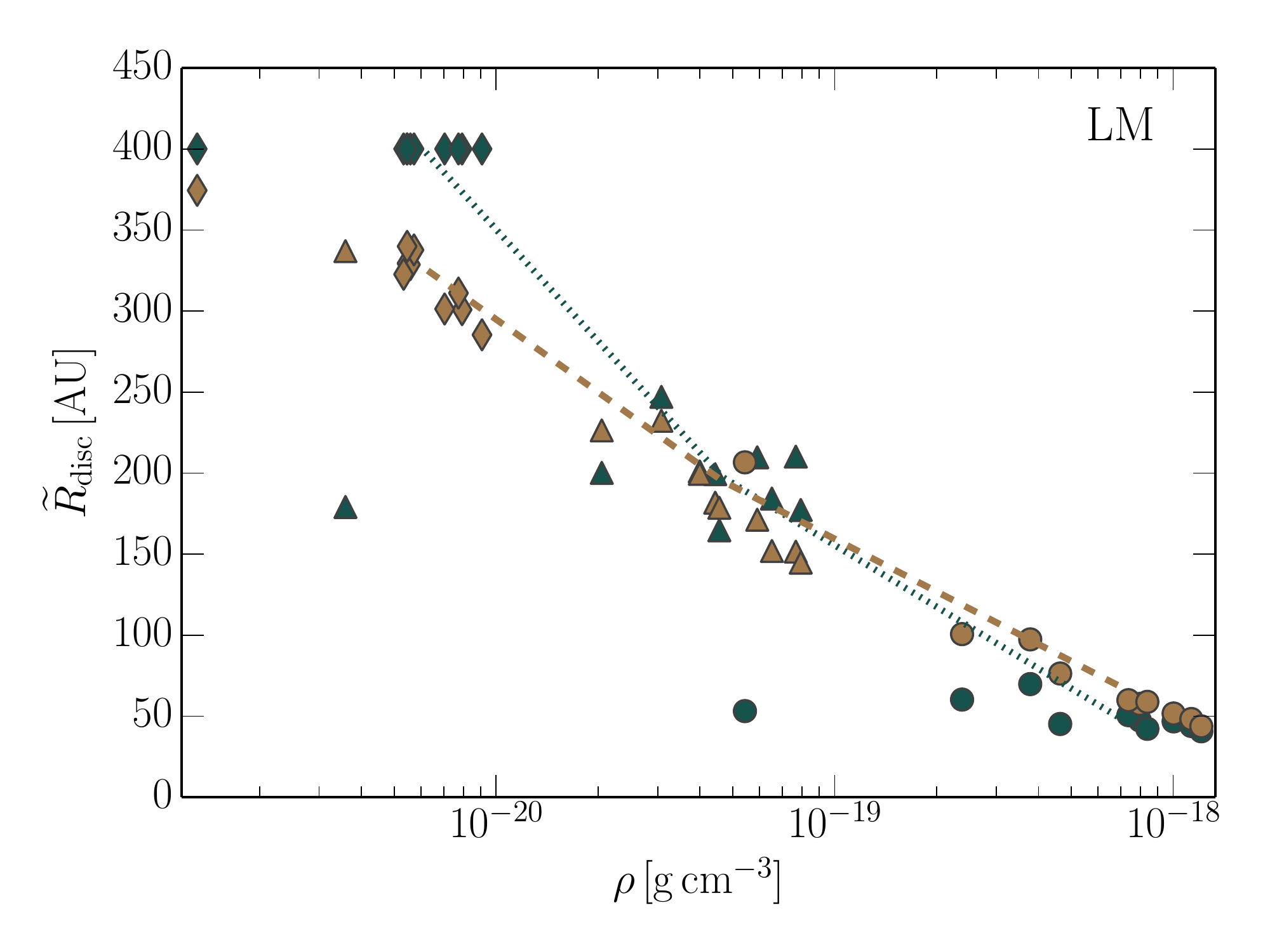}
        \subcaption{}
        \label{fig:300_density_application}
    \end{subfigure}
    \caption{\textbf{\emph{(a):}} The median disc radius at the end of the simulation plotted against the average stellar density within the cluster radius in the HM simulations. The colours correspond to the same process as in Fig.~\ref{fig:characterise_radius_application}, the symbols indicate the different cluster radii. The lines connect the averages of the ten simulations with the same parameters.\textbf{\emph{(b):}} The median disc radius at the end of the simulation plotted against the average gas density within the cluster radius in the LM simulations. The legend applies to both panels.}
\end{figure*}

In Fig.~\ref{fig:3000_stellar_density_application} we show the median disc radius as a function of the stellar density for our HM simulations with three different cluster radii. For a stellar mass fraction of 30\,\%, these simulations have the highest stellar density at each cluster radius. Fig.~\ref{fig:3000_stellar_density_application} shows that even at stellar densities $\ge 10^4 \,\mathrm{pc^{-3}}$, the process of face-on accretion leads to smaller disc sizes than dynamical encounters. As the stellar density increases with decreasing cluster radius, both the gas density and velocity dispersion increase as well, which enhances the effect of face-on accretion and ram pressure stripping. The highest stellar density in Fig.~\ref{fig:3000_stellar_density_application} corresponds to a gas density of $2 \times 10^{-17}\,\mathrm{g/cm^3}$. This is at the high end of what may be expected in the cores of star-forming regions if the star formation efficiency increases with gas density. These conditions illustrate that as long as the gas to stellar mass ratio is $\gtrsim 2$, face-on accretion is the dominant disc-truncation process regardless of the stellar density. Only by expelling gas or in case of more efficient star formation can dynamical encounters become the dominant process if the stellar density is sufficiently high. We find that for stellar densities $\lesssim 10^3$ pc$^{-3}$, dynamical encounters do not alter the disc radii, as was also shown in \citet{rosotti14} and \citet{vincke15}. 

In Fig.~\ref{fig:300_density_application} we show the results of our LM simulations with three different cluster radii. These simulations correspond to a stellar mass fraction of about 90\,\%. The scatter in the gas density between simulations with the same initial conditions is caused by the sampling of the initial mass function and the resulting variation in the remaining gas mass (as discussed in Sect.~\ref{sec:res_case_application}). The simulations with $R_{\rm cluster} = 0.25$\,pc can be interpreted as representing either the end of the embedded phase of the Trapezium cluster or an efficiently formed Trapezium cluster. In these Trapezium analogue simulations, labelled Trap90, the stellar density is $\approx 5 \times 10^3\,\mathrm{pc^{-3}}$. The contributions of face-on accretion and dynamical encounters to the decrease in disc radii are about equal in this case. In the most dense conditions, that is, $R_{\rm cluster} = 0.1$ and a stellar mass fraction of 90\,\%, dynamical encounters are mildly dominant, but the contributions of both processes are still very similar and severe:  $\widetilde{R}_{\rm disc} \lesssim 50\,$AU. If dynamical encounters result in a median disc radius of $\lesssim 80$ AU, they are very destructive and the parameterisation we use for their radii may no longer apply. The LM simulations with $R_{\rm cluster} = 0.5$\,pc in Fig.~\ref{fig:300_density_application} show that even at gas densities $\lesssim 10^{-20}\,\mathrm{g/cm^3}$ the process of face-on accretion is still relevant, because $\widetilde{R}_{\rm disc} = 327\,$ AU.

Our simulations show that the stellar density has to exceed a few times $10^3\,\mathrm{pc^{-3}}$ and simultaneously the stellar mass fraction has to be well above 30\,\% for dynamical encounters to be at least equally important as face-on accretion in the disc truncation process. Even if dynamical encounters are the dominant process, at stellar densities exceeding $10^4\,\mathrm{pc^{-3}}$ and a stellar mass fraction of 90\,\%, the influence of face-on accretion is still important. In these compact clusters, the corresponding velocity dispersion also enhances the effects of face-on accretion and ram pressure stripping. Generally, dynamical encounters become more important than face-on accretion in the regime where fly-bys are very destructive, resulting in $\widetilde{R}_{\rm disc} \lesssim 100 AU$.

\subsection{Disc masses}\label{sec:discmasses_application}

\begin{figure}[t]
\centering
    \includegraphics[width=0.49\textwidth]{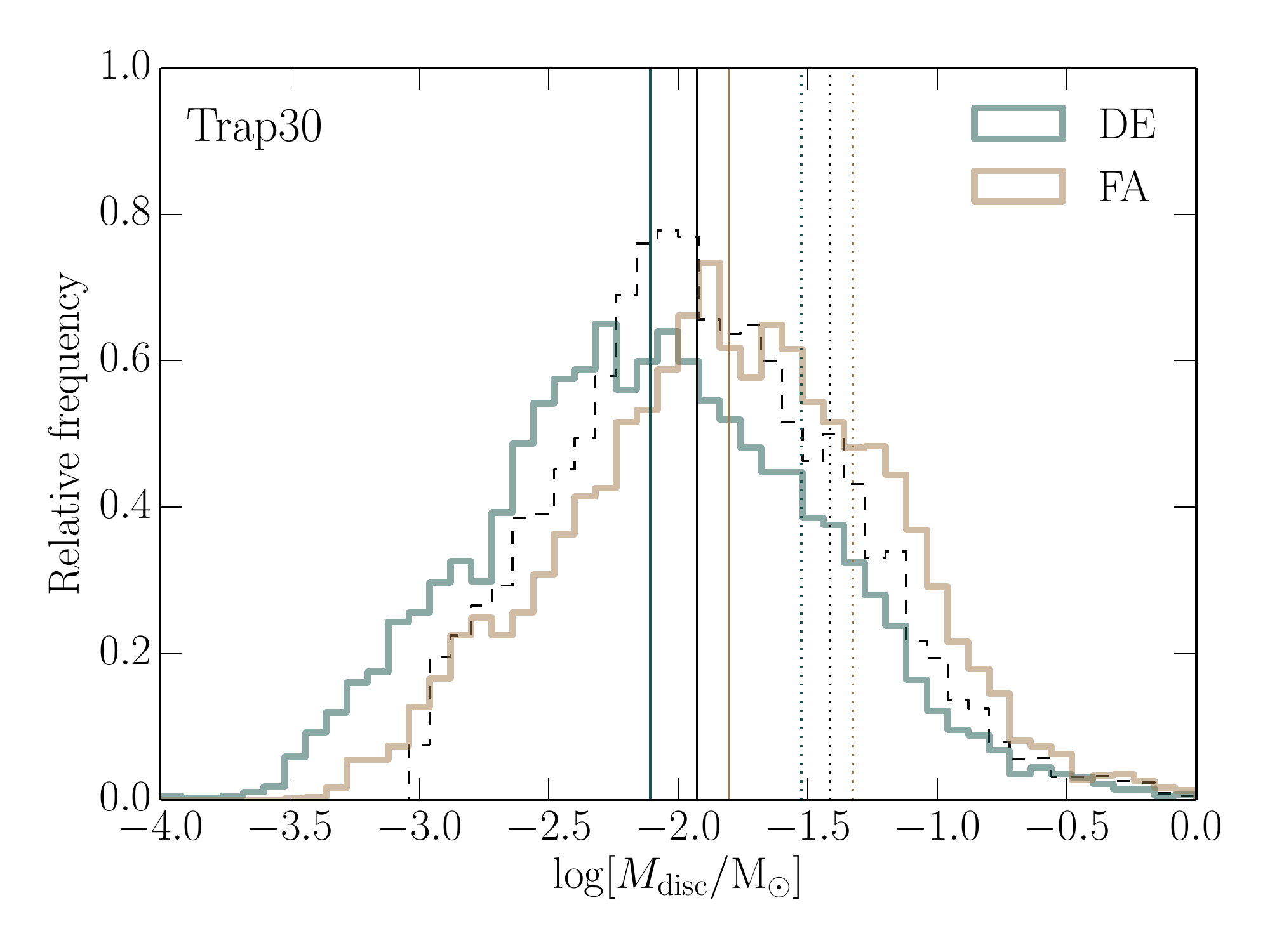}
    \caption{Relative distribution of disc masses at the end of our Trap30 simulations. Colours correspond to the previous figures. The initial disc mass distribution is shown in black dashed bins. The distribution is based on all the disc masses in the ten simulations we have performed with these parameters. The solid vertical lines in the corresponding colour indicate the median disc mass and the vertical dotted lines indicate the mean disc mass. \label{fig:discmass_distribution_application}}  
\end{figure}

In Fig.~\ref{fig:discmass_distribution_application} we show the distribution of disc masses in our Trap30 simulations. The net effect of a dynamical encounter is that the disc loses more mass than it accretes from the disc of the other star. In the process of face-on accretion, on the other hand, the discs generally gain mass as more mass is swept up than is stripped by the ram pressure. The distribution of disc masses resulting from dynamical encounters is therefore expected to be shifted to lower masses compared to the initial distribution and for face-on accretion the distribution should be shifted towards higher masses. This is demonstrated in Fig.~\ref{fig:discmass_distribution_application}. In all our simulations the median disc mass resulting from face-on accretion is higher than the median disc mass produced by dynamical encounters, as we discuss in Sect.~\ref{sec:res_masses_parameterspace_application}. As for the radius distribution, we choose the median disc mass, $\widetilde{M}_{\rm disc}$, to characterise the disc mass distribution. The median disc mass provides a better correspondence to the peak in the distribution than the mean disc mass, because the latter is determined by the high end of the disc mass spectrum.

\subsubsection{Disc masses as a function of the parameter space}\label{sec:res_masses_parameterspace_application}
\begin{figure*}[t]
    \begin{subfigure}[b]{0.5\textwidth}
        \centering
        \includegraphics[width=\textwidth]{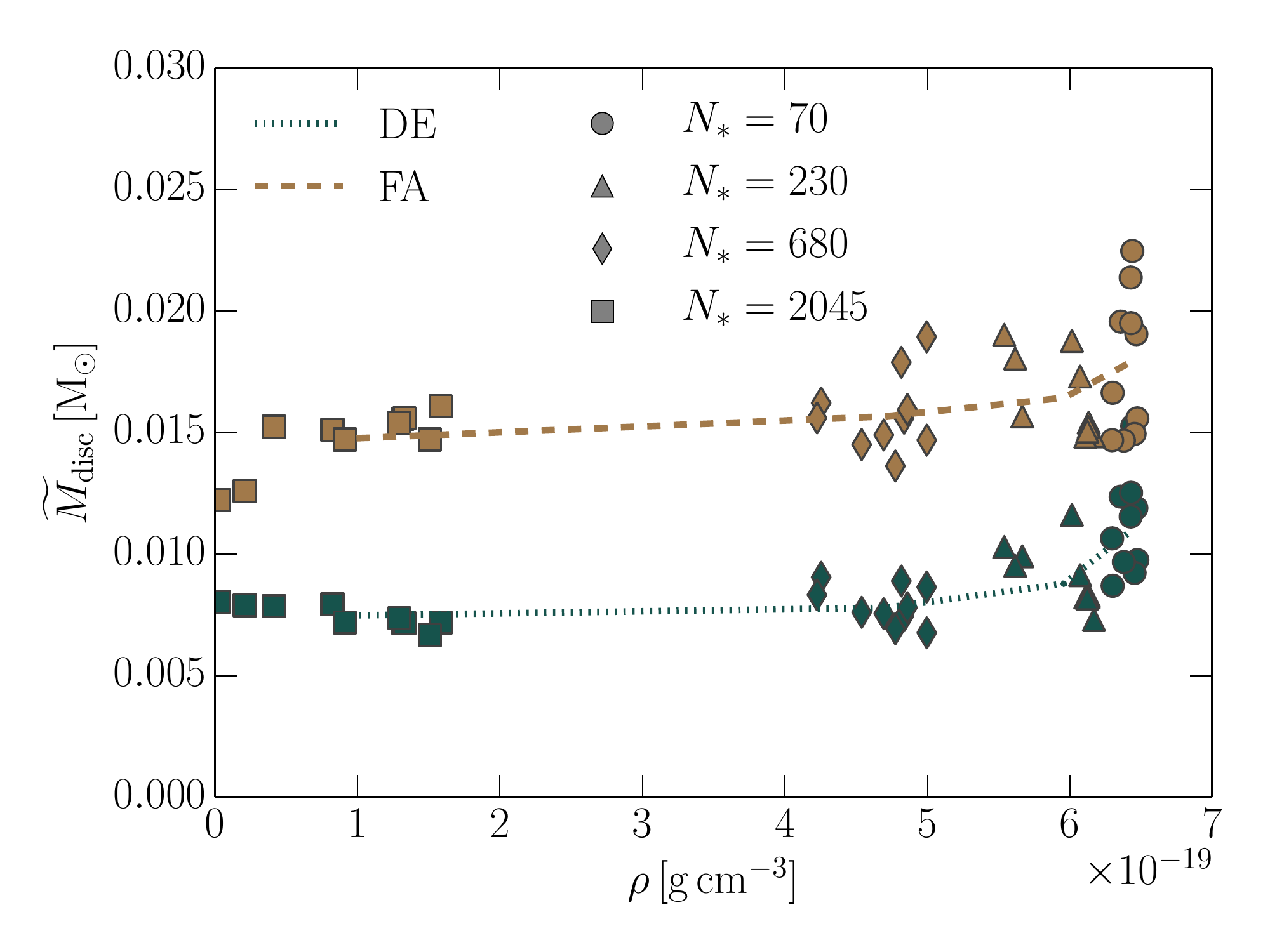}
        \subcaption{}
        \label{fig:discmass1000_density_application}
    \end{subfigure}
    \begin{subfigure}[b]{0.5\textwidth}
        \centering
        \includegraphics[width=\textwidth]{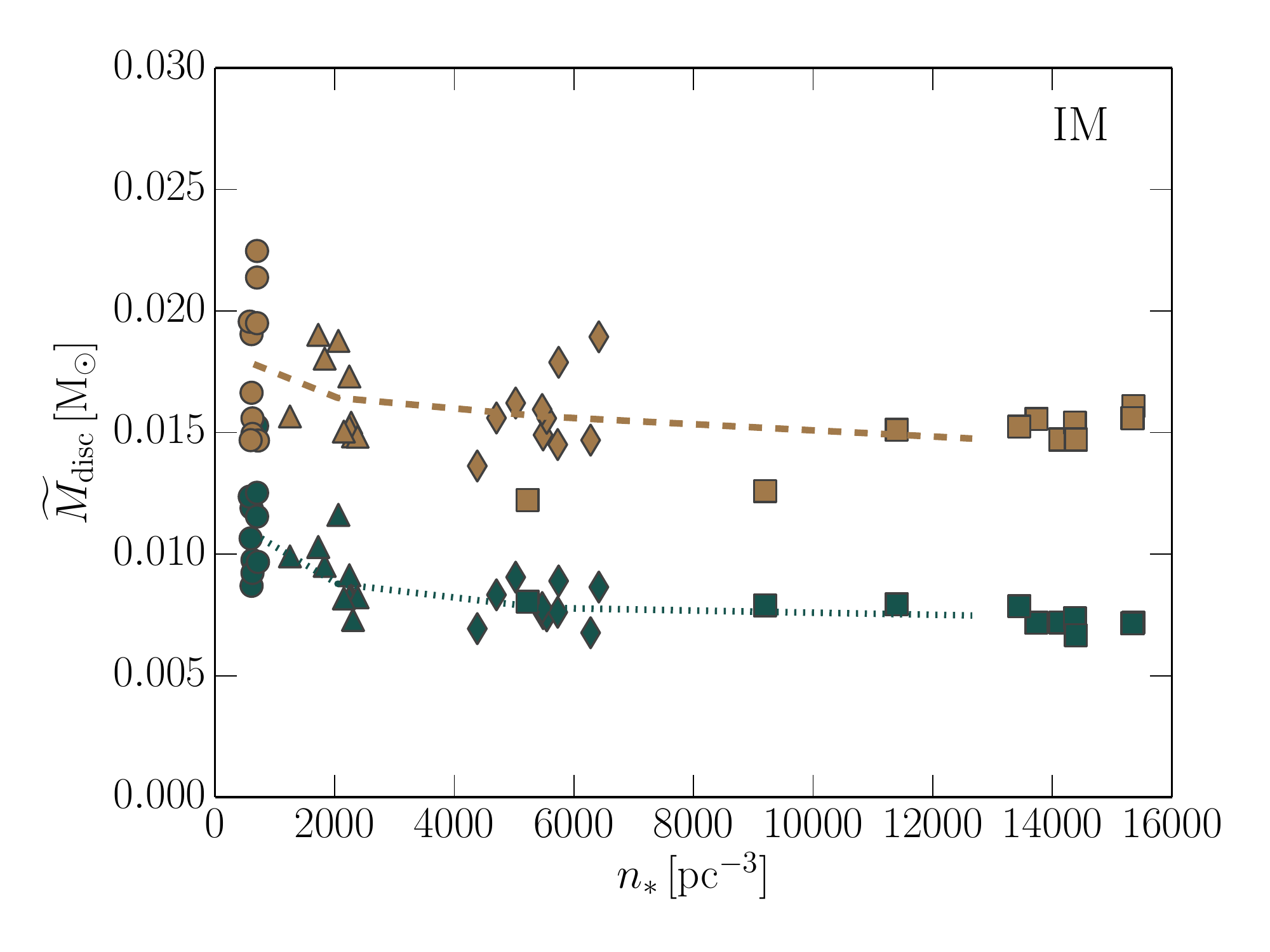}
        \subcaption{}
        \label{fig:discmass1000_stellar_density_application}
    \end{subfigure}
    \caption{Same as Figs.~\ref{fig:case_density_application} and \ref{fig:case_stellar_density_application} but here we show the median disc mass in the IM simulations.}
\end{figure*}

Figs.~\ref{fig:discmass1000_density_application} and \ref{fig:discmass1000_stellar_density_application} show the median disc mass as a function of the gas and stellar density, respectively, for our IM simulations. These results correspond to the median disc radii shown in Figs.~\ref{fig:case_density_application} and \ref{fig:case_stellar_density_application}. The median disc masses resulting from face-on accretion increase slightly with increasing gas density. Applying similar reasoning as for the median disc radius in Sect.~\ref{sec:res_case_application}, the median disc mass is expected to scale as $\widetilde{M}_{\rm disc} \propto \tau^{1/5} \propto \rho^{1/5}$, again assuming that averaged over the population only the mean gas density varies between IM simulations with a different number of stars. When the curve for the median disc masses in Fig.~\ref{fig:discmass1000_density_application} is fitted as a power-law of $\rho$, this gives an exponent of 0.07. The increase in the median disc mass with density is slower than theoretically expected, because ram pressure stripping is more effective at high gas density, removing mass from the discs and decreasing the median disc mass. For dynamical encounters, the median disc mass decreases with increasing stellar density, which is also expected theoretically. Yet the increasing stellar density has only a very small effect on the median disc mass once the stellar density $\gtrsim 5 \times 10^3$.

\begin{figure*}[ht]
    \begin{subfigure}[b]{0.5\textwidth}
        \centering
        \includegraphics[width=\textwidth]{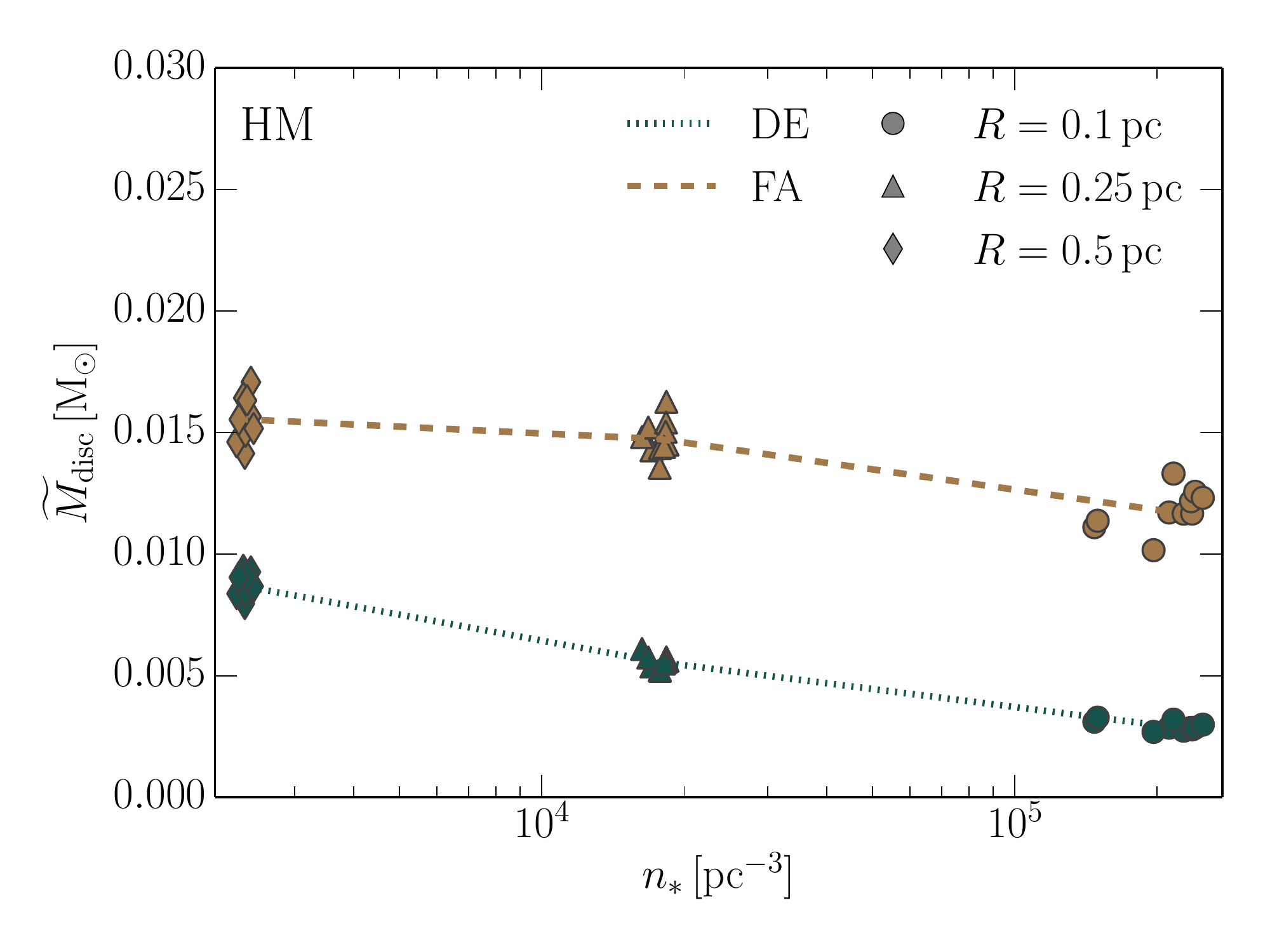}
        \subcaption{}
        \label{fig:discmass3000_stellar_density_application}
    \end{subfigure}
    \begin{subfigure}[b]{0.5\textwidth}
        \centering
        \includegraphics[width=\textwidth]{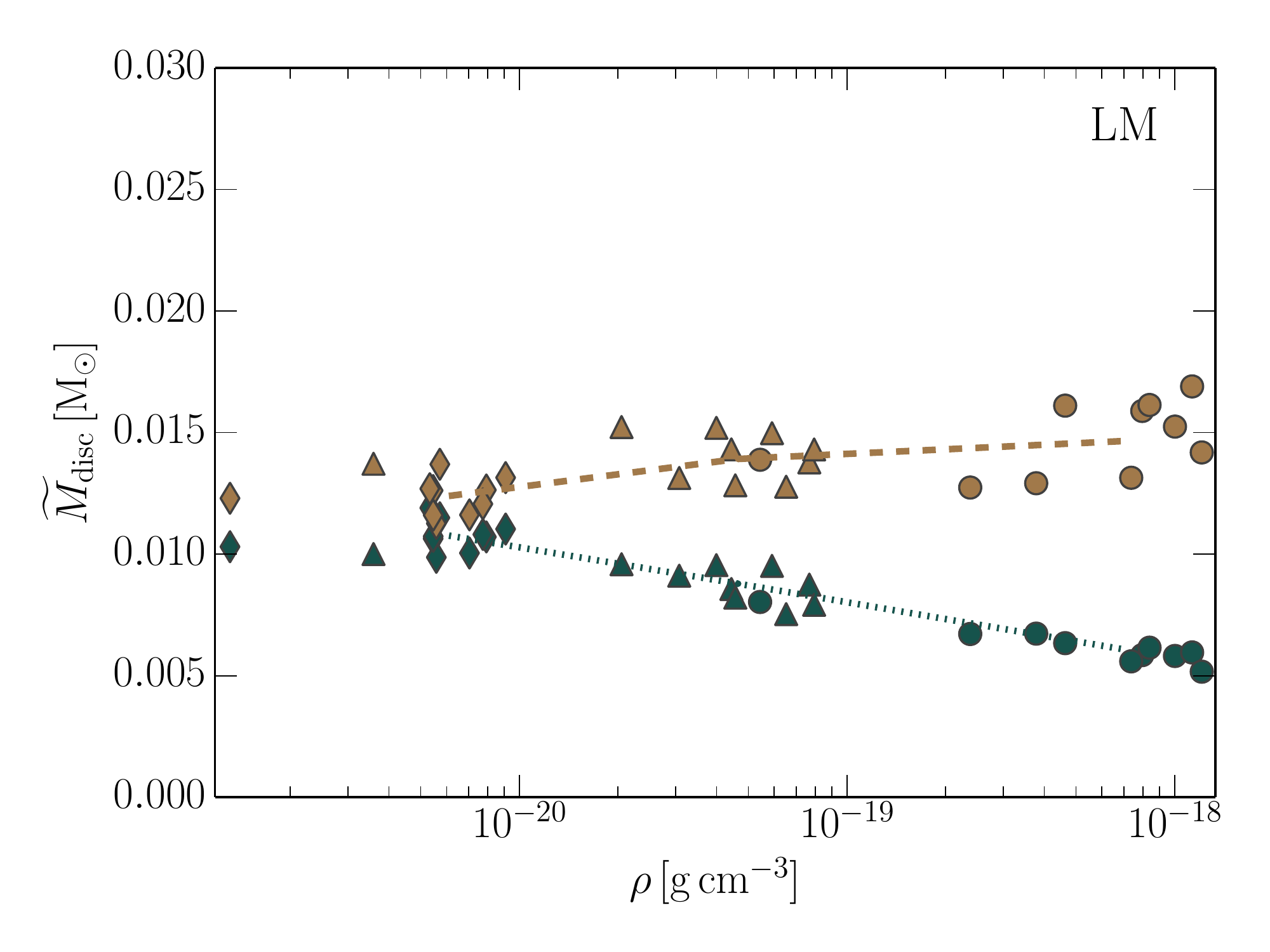}
        \subcaption{}
        \label{fig:discmass300_density_application}
    \end{subfigure}
    \caption{Same as Figs.~\ref{fig:3000_stellar_density_application} and \ref{fig:300_density_application} but here we show the median disc mass in the \textbf{\emph{(a)}} HM and \textbf{\emph{(b)}} LM simulations.} 
\end{figure*}

Figs.~\ref{fig:discmass3000_stellar_density_application} and \ref{fig:discmass300_density_application} show the median disc masses for the same simulations as shown in Figs.~\ref{fig:3000_stellar_density_application} and \ref{fig:300_density_application}. These figures illustrate that for either process, the cluster parameters have a small influence on the median disc mass. In the HM simulations (Fig~\ref{fig:discmass3000_stellar_density_application}), the median disc mass decreases with decreasing cluster radius for both processes because both the ram pressure and stellar density increase as the cluster radius decreases. In the HM simulations with $R_{\rm cluster} = 0.1\,$pc, ram pressure stripping is so strong that the median disc mass resulting from face-on accretion is 1.5\,\% lower than the initial median disc mass. In this particular case, the mass loss due to ram pressure stripping cannot be compensated by the mass gain from face-on accretion. In the LM simulations in Fig.~\ref{fig:discmass300_density_application}, the ram pressure is much lower and the median disc mass resulting from face-on accretion increases for more compact clusters that have a higher density and velocity dispersion. 

In general, we find that dynamical encounters lead to disc masses that are lower than those predicted by face-on accretion. Face-on accretion produces compact discs with a high surface density, while dynamical encounters generally result in larger and less massive discs.

\subsection{Virial ratios}
\begin{figure*}[ht]
    \begin{subfigure}[b]{0.5\textwidth}
        \centering
        \includegraphics[width=\textwidth]{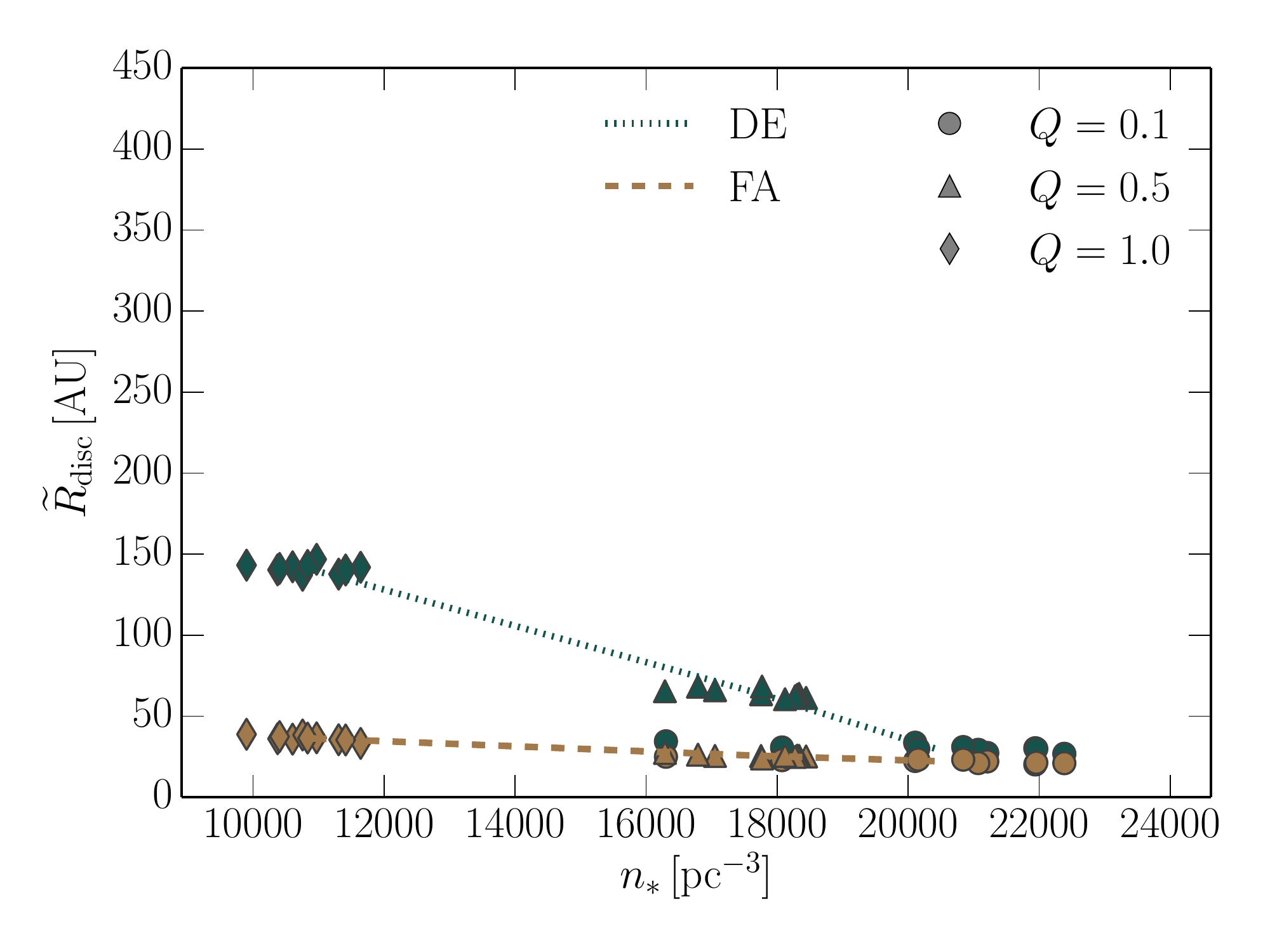}
        \subcaption{}
        \label{fig:radiiQ_application}
    \end{subfigure}
    \begin{subfigure}[b]{0.5\textwidth}
        \centering
        \includegraphics[width=\textwidth]{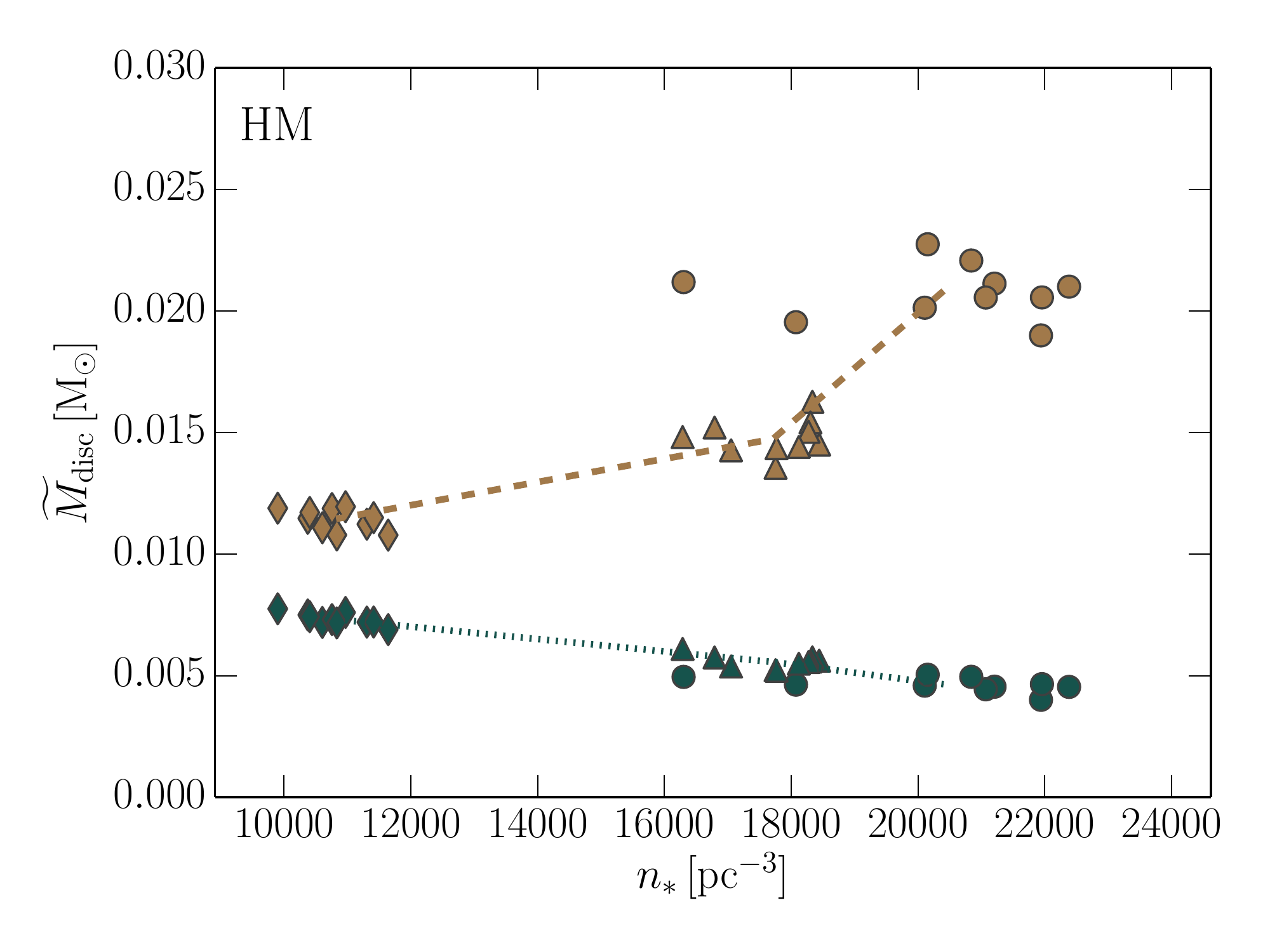}
        \subcaption{}
        \label{fig:discmassQ_application}
    \end{subfigure}
    \caption{\textbf{\emph{(a):}} The median disc radius plotted against the stellar density within the cluster radius at the end of the HM simulations with $R_{\rm cluster} = 0.25$\,pc for different virial ratios. \textbf{\emph{(b):}} Same as in the left figure, but here we show the median disc masses for the same simulations. The legend applies to both panels.}\label{fig:viralratios_application} 
\end{figure*}

In the simulations discussed in the previous section, we assume that the stars and gas are initially in virial equilibrium. We also performed simulations with `cold', that is, $Q_{\rm vir} = 0.1$, and `warm', $Q_{\rm vir} = 1$, initial conditions. The simulations that start dynamically cold go through a phase of contraction. This increases the stellar density and velocity dispersion. On the other hand, the simulations that start dynamically warm expand from the beginning, which causes the stellar density to decrease. In Fig.~\ref{fig:viralratios_application} we show the mean disc radii and masses in our HM simulations with $R_{\rm cluster} = 0.25$\,pc for the three virial ratios. For $Q_{\rm vir} = 1$ the stellar density is substantially lower at the end of the simulations and as a consequence the median disc radius resulting from dynamical encounters is larger than in the simulations in virial equilibrium. A virial ratio of $Q_{\rm vir} = 0.1$ leads to a higher stellar density at the end of the simulation and hence the median disc radius due to dynamical encounters is smaller. The median disc mass for dynamical encounters follows the same trend as a function of virial ratio. 

For face-on accretion, the median disc radius does not depend strongly on the virial ratio but there is a slight decreasing trend with decreasing virial ratio, that is, with increasing velocity dispersion. Simultaneously, the median disc mass increases with decreasing virial ratio. If the cluster contracts, the discs generally move through regions with higher density with a higher velocity compared to a cluster that expands. The discs are therefore able to sweep up more gas when the initial conditions are cold.

In essence, if the initial conditions are supervirial ($Q_{\rm vir} = 1)$, then face-on accretion is the dominant process in truncating the discs regardless of the other cluster characteristics we use in this work. Face-on accretion is also the dominant process if the initial conditions are subvirial and the stellar mass fraction is $\le 30$\,\%. Dynamical encounters become dominant if the stellar mass fraction is well above 30\,\% and the cluster goes through a phase of contraction.

\section{Discussion}

\subsection{Assumptions on accretion and mass loss}\label{sec:dis_assumptions_application}
We kept the mass of the stars and the gas potential constant in our simulations. Here we discuss the consistency of our simulations with this and our other assumptions on accretion and mass loss.

We assume that the total gas mass in the cluster remains constant in our simulations. In most simulations the net effect of face-on accretion does not change the total gas mass by $\gtrsim 1$\,\%, except in the simulations with the highest stellar-mass fraction. At most 20.9\,\% of the gas mass is accreted in one of these simulations (with $M_{\rm tot} = 3000\,\Msun$, $R_{\rm cluster} = 0.1$\,pc, $N_* = 6135$ and $Q_{\rm vir} = 0.5$). Dynamical encounters are the dominant cause of disc truncation in this regime, even if the effect of face-on accretion may have been overestimated due to the presumed constant gas density distribution. We therefore conclude that our results are not affected by the assumption that the amount of gas remains constant in our simulations.

Furthermore, the masses of the particles in the $N$-body simulation remain constant during the simulation. Except for the most compact simulations with the lowest stellar mass fraction, the fraction of stars, including their discs, that gain more than 10\,\% in mass is always less than 1\,\%  in our simulations that start in virial equilibrium. This small fraction of stars is similar to what is found by \citet{throop08} who simulated Bondi-Hoyle accretion for 4 Myr with similar cluster parameters and a star formation efficiency of 33\,\%. In the compact virialised simulations with the lowest stellar mass fraction, at most 11.6\,\% of the stars gained more than 10\,\% in mass in the simulations with $M_{\rm tot} = 1000\,\Msun$, $R_{\rm cluster} = 0.1$\,pc, $N_* = 70$ and $Q_{\rm vir} = 0.5$. This fraction increases to a maximum of 38.7\,\% in the initially subvirial simulation with $M_{\rm tot} = 300\,\Msun$, $R_{\rm cluster} = 0.1$\,pc, $N_* = 20$. A stellar mass fraction of roughly 3\,\% is probably not realistic in such dense conditions \citep[e.g.][]{bonnell03, burkert13} and we have therefore not discussed these simulations in detail. We merely use them to investigate how either process scales as a function of the initial conditions. 

We do not account for Bondi-Hoyle accretion in our simulations. Previous studies have modelled Bondi-Hoyle accretion onto stars in clusters \citep[e.g.][]{throop08, scicluna14, ballesteros-paredes15}. They find that the amount of accreted material scales with $M_*^2$ and for solar-mass stars it is typically not more than a few per cent of their initial mass, depending on the gas density and velocity dispersion. If the Bondi-Hoyle radius of a star is larger than its disc radius, the gravitational focusing of gas by the star enhances the effective surface area for accretion onto the disc. This increases the amount of gas that is swept up by the disc. We tested this effect on our face-on accretion model by correcting the mass flux for the gravitational focusing if the Bondi-Hoyle radius is larger than the disc radius \citep{bisnovatyi-kogan79, edgar04} but this did not change our distribution of disc radii and masses. Generally, the Bondi-Hoyle radius is smaller than the disc radius in our simulations. In case the Bondi-Hoyle radius is larger, the disc initially contracts faster and the net effect on the disc radius and mass is similar after 1 Myr. 

The amount of mass lost in a dynamical encounter is probably overestimated by Eq.~\ref{eq:dm_encounter_application}. \citet{breslau14} show that the surface density profile is affected by a dynamical encounter. Their Fig.~1b shows that the surface density within the cut-off radius can become higher than the initial surface density profile. Unfortunately, the effect of dynamical encounters on the surface density profile cannot be easily parameterised. For encounters between debris discs, \citet{jilkova16} showed that the minimum disc radius beyond which particles can be unbound, $R_{\rm unb}$, is proportional to the encounter pericenter and the mass of the accreting star divided by the total mass of both stars, in case of low-inclination and low-eccentricity encounters. However, not all material beyond this radius is lost from the disc. Their parameterisation for $R_{\rm unb}$ is similar to what we use for the new disc radius in Eq.~\ref{eq:trunc_encounter_application}. The redundant mass lost from the disc in our simulations is partly compensated by accretion of material that is lost from the other disc in the encounter. As long as there is net mass loss during an encounter, the relative influence of dynamical encounters and face-on accretion on the median disc mass does not change in our simulations.

\subsection{Viscous forces}
The viscous spreading of a protoplanetary disc makes it more vulnerable to dynamical encounters and ram pressure stripping. When a disc moves through an ambient medium, the disc size is determined by the process of face-on accretion and ram pressure stripping at the outer edge, that is, the disc cannot spread beyond a certain radius \citep{wijnen16, wijnen17}. For face-on accretion, neglecting viscous spreading is therefore a fair assumption, as long as the accretion time scale is shorter than the viscous time scale. We verified that for the accretion time scales in our simulations, this criterion is satisfied. In case of dynamical encounters, the increasing cross-section of the disc resulting from viscous spreading would increase the probability of a truncating encounter. However, during the embedded phase, the viscous spreading would be inhibited as outlined above. At the end of the embedded phase, viscous spreading may become important and might increase the probability of dynamical encounters. Our simulations with a stellar mass fraction of 90\,\% and stellar densities $\gtrsim 10^4\,\mathrm{pc}^{-3}$ show that, by that time, the process of dynamical encounters is dominant, without accounting for the viscous spreading. Including viscous spreading at this stage may ease the constraint on the stellar density. However, initial results by Concha Ram\'irez \& Vaher (in prep.) find that gas expulsion leaves the cluster in a supervirial state and including viscous spreading of the disc \citep[according to the self-similarity solutions of][]{lynden-bell74} does not increase the number of truncations in an expanding cluster.

\subsection{Drag forces}
As the discs move through an ambient medium they experience a drag force exerted by the ram pressure, which we have not taken into account. The part of the disc that is tightly bound to the star, which is determined by $R_{\rm tr}$ in Eq.~\ref{eq:Rtrunc_application}, can decelerate the star. We can estimate the time scale on which the drag force operates as $\tau_{\rm drag} = v/(dv/dt) \propto M_*/ \rho v R_{\rm disc}^2$. For an $0.4\,\Msun$ star in our Trap30 simulation, assuming an average disc radius of 200 AU, this corresponds to a velocity change of 5\,\% after 1 Myr. This time scale decreases with stellar mass but less massive discs contract faster. The average effect on the population is probably not larger than this 5\,\%, also considering that the disc of an $0.4\,\Msun$ is expected to be $\lesssim 200$ AU within 0.2 Myr in the centre of the Trap30 simulation. 

\subsection{Initial disc masses and radii}\label{sec:dis_discmasses_application}
We have assumed relatively high disc masses of $0.1 M_*$. The masses of protoplanetary discs are observed to be of the order of $0.01 M_*$ \citep{andrews05, andrews07}. These estimates may be off by an order of magnitude but, in general, disc masses are probably not much higher than $0.1 M_*$. Assuming lower disc masses in our simulations would not affect the relative influence of dynamical encounters on disc sizes in this study (cf. Eq.~\ref{eq:trunc_encounter_application}). A similar reasoning applies to the initial disc radii: Although the probability of experiencing a dynamical encounter depends on the initial disc radius, the resulting disc radius does not. Assuming initially larger disc radii would affect the tail of our distribution but not the median disc radius \citep[e.g. Fig. 2 in][]{vincke15}.

In case of face-on accretion, distributing a lower disc mass over the same surface area, or the same disc mass over a larger surface area, decreases the surface density of the disc. As a result, the disc is more vulnerable to ram pressure stripping and the disc contracts faster. If all other conditions remain the same, then over long time scales the disc radius scales as $\tau^{-2/5} \propto \Sigma_0^{2/5} \propto M_0^{2/5} R_0^{-4/5}$, where $M_0$ and $R_0$ are the initial disc mass and radius (Sect.~\ref{sec:faceontheory_application}). For a given disc mass, starting with an initial radius of 1000 AU instead of 400 AU thus causes the disc to contract about twice as fast. The same reasoning applies to assuming lower disc masses, except that the scaling with disc mass in the above reasoning is less strong. The truncation radius due to ram pressure stripping also decreases with decreasing initial surface density, although the dependence is weaker. The relative importance of face-on accretion with respect to dynamical encounters would therefore increase if the initial disc radii were larger and/or disc masses were smaller.

\subsection{Photoevaporation and winds}

As discussed in the introduction, neglecting the effect of external photoevaporation on the discs on a time scale of 1 Myr in an embedded cluster is probably a fair assumption. Winds from massive stars also affect the ambient gas density. They can locally clear the gas and create low-density cavities. The density in these regions can be orders of magnitude lower than the average gas density in the cluster. The wind velocity is three orders of magnitude higher than the velocity dispersion of the clusters in our simulations. Although the integrated mass flux on the disc would be similar, the high temperature difference between these ejecta and the disc may hamper efficient accretion. Two-dimensional simulations of supernova ejecta interacting with a protoplanetary disc show that only 1\,\% of the hot gaseous ejecta is intercepted by the disc, contrary to cold gas and dust that is accreted very efficiently onto a protoplanetary disc \citep{ouellette07, ouellette09, ouellette10, wijnen16, wijnen17}. Hence the integrated mass flux inside a wind cavity could be smaller by a factor of 100. The time scale on which these winds create cavities depends on the most massive star in the cluster and thus on the realisation of the initial mass function. Simulations of embedded clusters with $N_* = 1000$ that take feedback into account find that the impact of winds from massive stars on the gas distribution becomes relevant on a time scale of 1 Myr \citep{pelupessy12}, which is the duration of our simulations. The same authors find that supernovae are not expected to play a role until an age of roughly 10 Myr. The contraction of the disc caused by face-on accretion is faster in the beginning when stellar feedback may not have affected the gas distribution substantially yet. The winds could also strip the discs from their outer regions because of their high velocities.

We also neglect the influence of photoevaporation by the host star on both processes. Dynamical encounters are not affected by this process but the photoevaporative winds from the disc and jets may influence the face-on accretion process. Interaction between inflowing ambient gas and material from the star itself or evaporated disc material could slow down face-on accretion onto the disc. The winds from the star and the disc depend on the mass of the star and its spectral energy distribution. Extreme-ultraviolet radiation irradiates the inner few AU of the protoplanetary disc, while X-rays and far-ultraviolet radiation induce photoevaporation at disc radii of tens of AU and $\gtrsim 100$ AU, respectively \citep[e.g.][]{armitage11,williams11}. Irradiation by the central star may become dominant when the disc has shrunk to radii $\lesssim 100$ AU. Photoevaporation can further erode the disc from the inside out on time scales of $10^5$ yr after disc lifetimes of several Myr \citep[we refer to e.g.][ and references therein]{alexander14}. The increased surface density due to face-on accretion could make the disc more resistant against this erosion. Currently, these processes cannot be included in our models but they should be accounted for in future studies.

\subsection{Observational constraints}\label{sec:dis_observations_application}
\begin{figure}[t]
\centering
    \includegraphics[width=0.49\textwidth]{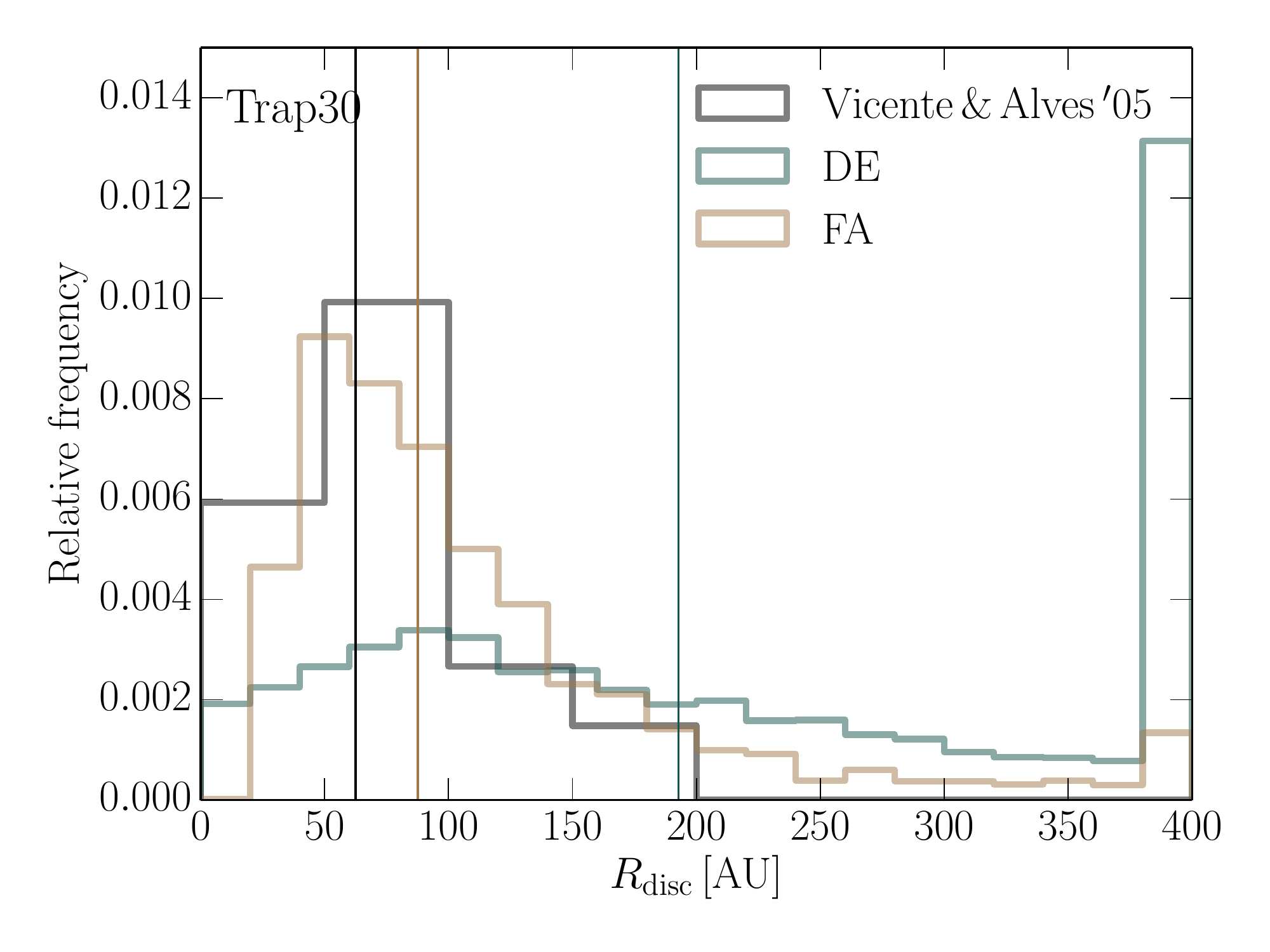}
    \caption{Identical to Fig.~\ref{fig:hist_application} but here we exclude all stars with mass $\le 0.1\, \Msun$ and compare to the observational sample of the ONC from \citet{vicente05}. \label{fig:discradii_obs_application}}  
\end{figure}
\noindent
The process of dynamical encounters predicts a larger spread in observed disc radii than face-on accretion. For example, in the Trap30 simulations, the radius distribution resulting from face-on accretion is concentrated around a clear peak (cf. Fig.~\ref{fig:hist_application}), whereas the distribution produced by dynamical encounters is much broader. In Fig.~\ref{fig:discradii_obs_application} we compare the radius distributions of the Trap30 simulations to observations of disc radii in the ONC. \citet{vicente05} assembled a sample of 149 protoplanetary discs out of 300 young stellar objects, which they claim is complete down to a disc radius of 50 AU. To compare to the observations, we only show the simulation results for stars with a mass $> 0.1\,\Msun$, which roughly corresponds to the latest spectral type of the stars to which the discs could be matched by \citet{vicente05}. The distribution resulting from face-on accretion is in better agreement with the observed distributions. Based on the resolution limit of their data, \citet{vicente05} estimate that 40 to 45 \,\% of the discs in the Trapezium cluster have radii larger than 50 AU. This agrees with what we find for the whole simulated population in Fig.~\ref{fig:hist_application}. Excluding the very low-mass stars ($\le 0.1\,\Msun$) increases the median disc radius for both processes by roughly 40 AU, because the discs of these stars generally have small radii. This is also demonstrated by the absence of the bin with the smallest radii for face-on accretion in Fig.~\ref{fig:discradii_obs_application}, when compared to Fig.~\ref{fig:hist_application}. The relative influence of both processes is not affected by the inclusion or exclusion of these stars in our analysis. 

The clusters simulated in this work are generally more dense than the observed band of young stellar clusters in the mass-radius plane \citep[e.g.][]{pfalzner16} and $N_*$-radius plane \citep[e.g.][]{kuhn15}. We have shown in this work that lowering the stellar and gas density consistently, that is, decreasing the cluster mass for a given radius or increasing the cluster radius for a given mass, increases the influence of face-on accretion relative to dynamical encounters. The effect of face-on accretion can therefore be expected to be even stronger with respect to dynamical encounters in the parameter space covered by the majority of observed embedded clusters.

The initial conditions in this work are idealised but our simulations illustrate that the relative importance of both processes can be constrained by observations of disc radii. Furthermore, the combination of such observations with observations of the stellar and ambient gas density might put constraints on the duration of the embedded phase of clusters and their star formation efficiencies. A comparison based on disc masses is less straightforward. As discussed in Sect.~\ref{sec:dis_discmasses_application}, disc masses can only be estimated from observations to within a factor of 10. According to our simulations this is not enough to distinguish between either process based on the resulting disc masses.

\subsection{Combined effect of dynamical encounters and face-on accretion}

The prescriptions for dynamical encounters and face-on accretion cannot be coupled consistently, because it is not evident how the mass that is accreted in a dynamical encounter should be accounted for in the face-on accretion model. We performed test simulations that included the combined effect of both processes by neglecting the mass accreted in a dynamical encounter. These simulations show that when face-on accretion is the dominant process, the additional truncation caused by dynamical encounters on the disc radius is marginal, unless the influence of both processes is comparable. On the other hand, if dynamical encounters are the dominant truncation mechanism then face-on accretion further decreases the disc size. 

The median disc mass that results from combining both processes is always intermediate between the median disc masses predicted by either process independently, whether the mass that is accreted in a dynamical encounter is added at the end of the simulation or not. The value of the median disc mass resulting from the combined effect of both processes leans towards the value predicted by the process that dominates the truncation of the disc radii.  

\section{Conclusions}
By including the processes of face-on accretion and dynamical encounters in $N$-body simulations of embedded clusters, we find that face-on accretion, including the effect of ram pressure stripping, is dominant in truncating protoplanetary discs if the fraction of mass in stars is $\lesssim 30\,\%$, regardless of other cluster parameters. The stellar mass fraction has to be well above 30\,\% and simultaneously the stellar density has to exceed a few times $10^3$ pc$^{-3}$ in order for dynamical encounters to have a comparable effect on the truncation of the discs. Even at stellar densities $\gtrsim 10^4$ pc$^{-3}$ and for gas mass fractions $\lesssim 10\,\%$, face-on accretion has a comparable effect to dynamical encounters, but in these circumstances both processes are destructive, resulting in median disc radii $\lesssim 100$ AU. We confirm the results from previous studies that dynamical encounters require stellar densities $\gtrsim 10^3$ pc$^{-3}$ to be effective at all \citep{rosotti14, vincke15}. This implies that dynamical encounters only become relevant compared to face-on accretion either (1) in clusters with stellar densities $\gtrsim 10^3$ pc$^{-3}$ and extremely high stellar mass fractions, $\gtrsim 90\,\%$ or (2) at the end of the embedded phase of the cluster if the stellar density is similar and discs have not yet shrunk sufficiently due to face-on accretion. Our modelling of the face-on accretion process does not take viscous spreading of the disc or photoevaporative winds from the star and disc into account, which may both reduce the contraction of the disc due to face-on accretion. On the other hand, we have assumed massive discs with $M_{\rm disc}=0.1\,M_*$, while lower-mass discs that are subject to face-on accretion contract faster and are more vulnerable to ram pressure stripping. Of the available prescriptions for the effect of dynamical encounters on the disc radius, we use the one that results in the smallest radius.

Face-on accretion leads to discs that are compact and have a relatively high surface density. On the other hand, dynamical encounters generally result in larger and less massive discs.

\begin{acknowledgements}
We are thankful to Lucie J\'ilkov\'a, Francisca Concha Ram\'irez, Eero Vaher and Vincent H\'enault-Brunet for valuable discussions. We also thank the referee for his/her comments.
This research is funded by the Netherlands Organisation for Scientific Research (NWO) under grant 614.001.202.
\end{acknowledgements}

\bibliographystyle{aa} 
\bibliography{phdbib}

\end{document}